\documentclass[11pt,a4paper]{article}
\pdfoutput=1
\usepackage{jheppub1}
\usepackage{rotating}
\usepackage[bb=boondox]{mathalfa}
\usepackage{wrapfig}
\usepackage{comment}
\usepackage{graphicx,graphbox}
\usepackage{amsfonts}
\usepackage{amsmath}
\usepackage{slashed}
\usepackage{amssymb}
\usepackage{latexsym}
\usepackage{multirow}
\usepackage{hyperref}
\usepackage{enumitem}
\hypersetup{colorlinks=true,citecolor=red,linkcolor=magenta,urlcolor=blue}
\usepackage{relsize}
\usepackage{booktabs}
\usepackage{subfigure}
\usepackage{cleveref}
\usepackage{fancyhdr}
\usepackage{float}
\usepackage{mmacells}
\usepackage{graphicx}
\usepackage{microtype}
\usepackage{tabularx}
\usepackage{xcolor}
\usepackage{siunitx,adjustbox,bm}
\usepackage[bottom]{footmisc}
\usepackage{makecell}
\usepackage[normalem]{ulem}

\usepackage{mathtools}
\usepackage[thinc]{esdiff}
\usepackage[euler]{textgreek}
\usepackage{braket}
\usepackage{physics}
\usepackage{xfrac}
\usepackage{soul}
\usepackage{stmaryrd}
\usepackage{pifont}
\newcommand{\mathematicanb}{\href{https://github.com/shakeel-hep/HK-Effective-Action}{\underline{Mathematica notebook}}}

\newcommand{\nn}{\nonumber}

\renewcommand{\tr}{\text{tr}}
\newcommand{\eff}{\text{eff}}
\renewcommand{\L}{\mathcal{L}}
\renewcommand{\c}{\text{c}}
\newcommand{\C}{\mathcal{C}}
\renewcommand{\O}[1]{{\mathcal O \left({#1}\right)}}
\begin{document}
\allowdisplaybreaks
%%%%%%%%%%%%%%%%%%%%
\title{One-loop Effective Action up to any Mass-dimension for Non-degenerate Scalars and Fermions including Light-Heavy Mixing}
%%%%%%%%%%%%%%%%%%%%	
\abstract{We propose a Heat-Kernel-based method to compute one-loop effective action up to any mass-dimension with arbitrary numbers of non-degenerate scalars and fermions. We demonstrate our prescription by computing the dimension six effective action for arbitrary numbers of non-degenerate scalars, which are consistent with degenerate results presented in prior studies. We have further shown that the effective action for any partially degenerate mass-spectrum can be easily derived from that for the non-degenerate case by employing suitable limits for masses. Our prescription allows us to express the effective action in terms of local operators for the scenarios that involve light-heavy mixing as well, simply by employing an infrared regulator in the non-degenerate action. Along with the formalism, we have also provided the algorithm and outlined the  Mathematica program to compute the result up to any mass dimension.}	
%%%%%%%%%%%%%%%%%%%%%%%%%%%%%%%%%%%%%%%%%%%%%%%%%%%%%%%%%%%%%%%%%
\author[a]{Upalaparna Banerjee,}
\author[a]{Joydeep Chakrabortty,}
\author[b]{Shakeel Ur Rahaman,}
\author[a]{Kaanapuli Ramkumar}

\emailAdd{upalab@iitk.ac.in, joydeep@iitk.ac.in, kaanapuliramkumar08@gmail.com}
\emailAdd{shakeel.u.rahaman@durham.ac.uk}
%%%%%
\affiliation[a]{Indian Institute of Technology Kanpur, Kalyanpur, Kanpur 208016, Uttar Pradesh, India.}
\affiliation[b]{Institute for Particle Physics Phenomenology, Department of Physics, Durham University, Durham DH1 3LE,
U.K.}

%%%%%%%%%%%%%%%%%%%%%%%%%%%%%%%%%%%%%%%%%%%%%%%%%%%%%%%%%%%%%%%%%
	
%%%%%%%%%%%%%%%%%%%%%%%
\preprint{}
%%%%%%%%%%%%%%%%%%%%%%%
%\pacs{}
%%%%%%%%%%%%%%%%%%%%%%%

\maketitle
	
%%%%%%%%%%%%%%%%%%%%%%%%%%%%%%%%%%
\newpage
\section{Introduction}
%%%%%%%%%%%%%%%%%%%%%%%%%%%%%%%%%%
In the recent precision era of particle physics, as experimental measurements have reached unprecedented accuracy, the need for a theoretical framework that is capable of accommodating such high-precision data has grown. Thus, Effective Field Theory (EFT) \cite{Weinberg:1980wa,Georgi:1994qn,Manohar:2018aog,Cohen:2019wxr}, known for its characteristics of systematic improvement, provides a robust framework to ensure accurate predictions for the measurements in the low-energy experiments while also incorporating the new physics effects of the high-energy scale indirectly. In the regime of EFT, provisions exist for taking into account the impact of these high-scale phenomena, both in a model-independent (Bottom-Up) \cite{BUCHMULLER1986621,Grzadkowski:2010es,Lehman:2014jma,Murphy:2020rsh,Li:2020gnx,Li:2022tec,Banerjee:2020jun,Anisha:2019nzx,Banerjee:2019twi}, and model-specific (Top-Down) manner. 

Over the past few years, the higher mass-dimensional terms beyond dimension six in the effective expansion of the low-energy Lagrangian have captured significant interest from both theoretical and phenomenological sides for a number of reasons \cite{Dawson:2021xei,Hays:2018zze,Corbett:2021eux,DasBakshi:2023htx,DasBakshi:2022mwk,Chala:2021pll,Alioli:2020kez,Degrande:2013kka,Ellis:2020ljj,Hays:2020scx,Dawson:2023ebe,Dawson:2022cmu,Ellis:2019zex,Corbett:2023qtg,Ellis:2023zim,Degrande:2023iob,Banerjee:2023bzl}. The dimension eight operators can play a significant role in capturing the intricate signatures of the complete UV theories, mainly when the dimension six operators are not produced at tree-level. Thus including them in the computation can help improve the predictions for observables from a model-specific point-of-view. On the phenomenological side, the ongoing search process for the presence
of the neutral triple and quartic gauge couplings forces us to go beyond the dimension six level.  Besides, taking into account the contributions from one-loop matching can also be necessary to obtain accurate constraints on the BSM parameter space.

Consequently, streamlining the matching procedure that exists beyond the regime of tree-level and dimension six to bridge the gap between high-energy UV theories and low-energy effective theories has gained renewed attention. At the level of dimension six, this process is usually carried out through any of these two approaches: either by computing the correlation functions of the light fields in the UV theory and subsequently imposing the condition that one field is significantly heavier than others, thereby extracting the Wilson coefficients; or by adopting the functional approach that directly evaluates the path integral \cite{Gaillard:1985uh,Chan1985,Cheyette:1987qz}. The latter approach manifestly preserves the gauge covariance and can be applied without considering the specifics of the UV origin. Thus, the comprehensive form for the Universal One-Loop Effective Action (UOLEA) \cite{Henning:2014wua,Drozd:2015rsp,Henning:2016lyp,Ellis:2016enq,delAguila:2016zcb,Ellis:2017jns,Kramer:2019fwz,Angelescu:2020yzf,Ellis:2020ivx} was formed, first considering the effects of only heavy loops with multiple degenerate scalars, which was later generalized to include the heavy fields with non-degenerate mass-spectra and even the contributions from the loops containing scalars as well as fermions. Subsequently, a number of automated tools have been developed to aid the matching procedure \cite{Aebischer:2018bkb,Carmona:2021xtq,Criado:2017khh,Celis:2017hod,Bakshi:2018ics}. 

Until recently, the UOLEA was limited to dimension six. However, in light of the recent advancements in phenomenological studies that use EFT as their theoretical backbone and also extend to explore the impact of dimension eight operators, our previous works have introduced the comprehensive UOLEA up to dimension eight for integrating out both heavy scalars and fermions \cite{Banerjee:2023iiv,Chakrabortty:2023yke}. The complete and readily applicable one-loop effective actions, as presented in these works, were obtained by employing the approach of the Heat-Kernel (HK) \cite{Minakshisundaram:1949xg,Minakshisundaram:1953xh,Hadamard2003,DeWitt:1964mxt,Seeley:1969,Schwinger:1951nm,Vassilevich:2003xt,Avramidi:2001ns,Avramidi:2015ch5,Kirsten:2001wz,Fulling:1989nb}.  The perk of using this method is that instead of having to compute the Feynman diagrams and performing loop integrals or the covariant derivative expansion (CDE) \cite{Henning:2014wua,Henning:2016lyp}, the effective action operators of a particular mass dimension can be directly expressed in terms of the coefficients of the perturbative expansion of the HK. These Heat-Kernel Coefficients (HKCs) can be computed once and for all in terms of the parameters of the second-order differential operators in a systematic and algorithmic manner. Since the HKCs are just functional of the parameters of the elliptic second-order operator, the problem of integrating out fields of different spins, like scalars and fermions, can be unified, and the same HKCs can be used in both cases.

In our recent works on calculating the effective action operators up to mass-dimension eight by integrating out heavy scalars \cite{Banerjee:2023iiv} and fermions \cite{Chakrabortty:2023yke}, we restricted ourselves to the scenario of having only degenerate particles. This case is simpler to solve since the HKCs can be readily calculated from the De-Witt recursion relations \cite{Hadamard2003,Minakshisundaram:1953xh,DeWitt:1964mxt}. However, in most realistic scenarios, it is common to have fields that need to be integrated out, do not belong to the same multiplet, and thus have different masses \cite{Drozd:2015kva,Huo:2015exa}. For instance, in the context of constructing operators for the Low Energy Effective Theory (LEFT) \cite{Jenkins:2017jig,Dekens:2019ept,Liao:2020zyx}, one must integrate out heavy fields like top quark, Higgs as well as the massive gauge bosons. Other than that, while performing one-loop matching, it is essential to consider the effects of the heavy-light mixed loops as well. Hence, from a phenomenological perspective, it becomes imperative to construct a framework that can take care of the loop effects of non-degenerate heavy fields in the most general manner.

In this paper, we extend our previous formalism to treat the case of having non-degenerate particles in the loop and obtain the one-loop effective action that is applicable for both scalars and fermions. The degenerate formula can be obtained as a limiting case of the non-degenerate results and the heavy-light mixed loop contributions can also be computed naturally from this formalism. We have built a Mathematica program that can compute the effective operators up to any mass dimensions for any number of non-degenerate particles. This program also includes provisions to compute the heavy-light mixing contributions.

The paper is arranged in the following manner. Sec. \ref{sec:heat-kernel-intro} describes the methodology of extending the Heat-Kernel method to the non-degenerate particles for integrating out heavy scalars. Sec. \ref{sec:procedure} elaborates on the methodology described in Sec. \ref{sec:heat-kernel-intro}, and provides the guidelines to obtain one-loop effective action operators with two non-degenerate particles in the loop. The details of taking the degenerate limit and obtaining the heavy-light mixed loop contributions are also dealt with in this section. Following this, we briefly explain how this method can be used to compute the effective action operators by integrating out non-degenerate heavy fermions in the loop in Sec. \ref{subsec:heat-kernel-fermion}. In Sec. \ref{sec:code}, the outline of the program and details on how to use it is provided. Finally, we conclude this article in Sec. \ref{sec:conclusion}.

%%%%%%%%%%%%%%%%%%%%%%%%%%%%%%%%%%%%%%%%%%%%%%%%%%
\section{One-loop Effective Action for Scalars with Non-degenerate Mass Spectrum}
\label{sec:heat-kernel-intro}
%%%%%%%%%%%%%%%%%%%%%%%%%%%%%%%%%%%%%%%%%%%%%%%%%% 
We begin by considering the UV Lagrangian for a model composed of a set of scalars with non-degenerate mass-spectra,
\begin{eqnarray}
    \mathcal{L}_{\text{Scalars}} = \Phi_a^{\dagger}\,(D^2\delta_{ab}+M_a^2\delta_{ab}+ U_{ab})\, \Phi_{b}\;,
\end{eqnarray}
where $M_a$ is the mass of the $a$-th particle, and $U_{ab}$ denotes the field-dependent interaction between particles $a$ and $b$.

In order to perform one-loop matching after integrating out the heavy fields using the functional method, we, in general, rely on the background field method. We define the fields, that appear in the loop, as quantum fluctuations around their respective classical backgrounds as, e.g., $\Phi = \Phi^c + \phi$. Here, $\phi$ is the quantum fluctuations and the classical background $\Phi^c$ is the solution of the following equation $\left(\delta S/\delta \Phi\right)\vert_{\Phi=\Phi^c} = 0$. The effective action is then given by,
{\small\begin{align}\label{eq:effective-action}
        e^{i\,S_{\text{eff}}[\Phi^c]} = \int [\mathcal{D}\Phi^\dagger][\mathcal{D}\Phi]e^{i\,S[\Phi]}
        &= \int [\mathcal{D}\phi^\dagger][\mathcal{D}\phi] \exp\left[i\,\left(S[\Phi_c]\,+\,\frac{\delta^2 S}{\delta\Phi^{\dagger}\delta\Phi}\bigg\vert_{\Phi=\Phi^c}\,\phi^\dagger \phi\,+\,\mathcal{O}(\phi^3)\right)\right],\nn\\
        S_{\text{eff}} &\approx S[\Phi^c]\,+\,i\ln\left(\text{Det}\,\frac{\delta^2 S}{\delta\Phi^{\dagger}\delta\Phi}\bigg\vert_{\Phi=\Phi^c}\right).
\end{align}}

Hence, the one-loop effective action, from the terms that are quadratic in fluctuations, can be written in terms of the second-order differential operator as
\begin{equation}\label{eq:effective-action}
     S_{\text{eff,1-loop}} = i\,c_s\, \text{Tr} \log(D^2+M^2+ U)\;,
\end{equation}
where $c_s$ is $1/2$ for real scalars and $1$ for complex scalar fields, $M$ contains the masses of each field along its diagonal elements, and the elements of $U$ are given by $U_{ab}$.

At this point, using the following identity, $\ln\lambda=-\int_{0}^{\infty}\cfrac{dt}{t}\,e^{-t\lambda}$, we can express the `log' function in Eq.~\eqref{eq:effective-action}, in terms of the exponent of the second-order elliptic operator operator $\Delta = (D^2+M^2+U)$. This can be directly linked to the Heat-Kernel, $K(t,x,y,\Delta)$ corresponding to $\Delta$ and hence, the Euclideanised effective action can be written as \footnote{Since the second order differential operator in the action is a hyperbolic operator, in Minkowski space, to obtain the Heat-Kernel corresponding to $\Delta$, one needs to perform Wick rotation to express it in Euclidean space. For a detailed discussion see Ref~\cite{Banerjee:2023iiv,Chakrabortty:2023yke}. The Heat-Kernel corresponding to this strong elliptic operator then can be expressed as,
\begin{equation}
    K(t,x,y,\Delta) = \bra{y}e^{-t\Delta}\ket{x}, \quad t>0.
\end{equation}},  
\begin{equation}\label{eq:effective-action-with-HK}
    S_{\text{eff,1-loop}} = c_s\ \text{Tr}\int d^4 x\int_0^{\infty} \cfrac{dt}{t} \,K(t,x,y,\Delta) = c_s\ \text{tr}\int d^4 x\int_0^{\infty} \cfrac{dt}{t} \,K(t,x,x,\Delta)\;.
\end{equation}
The functional trace `Tr' in Eq.~\eqref{eq:effective-action-with-HK} acts over all coordinates, including space-time coordinates, therefore computing the Heat-Kernel at the coincidence point $x=y$ only leaves a trace `tr' over the internal indices.  

In our previous works \cite{Banerjee:2023iiv,Chakrabortty:2023yke} while considering degenerate masses, we expressed $K(t,x,x,\Delta)$ in Eq.~\eqref{eq:effective-action-with-HK}, as a product of free Heat-Kernel ($K_0=K(t,x,y,\partial^2+M^2)$) of a second-order elliptic operator without any interactions, and a power series expansion in $t$, around $t=0$, of the perturbative interactions with coefficients $[b_k]$, called the Heat-Kernel coefficients (HKC). These HKCs were then computed in an algorithmic manner using the De-Witt recursion relations. When dealing with multiple non-degenerate fields, the approach is slightly different \cite{Osipov:2001bj,Osipov:2021dhc}. In both cases the convergence of the integral over the Heat-Kernel parameter $t$, henceforth called the $t$-integral, is given by the mass term, $e^{-M^2t}$ of the Heat-Kernel. In the case of having particles with degenerate masses, it is straightforward to take out the mass term from the Heat-Kernel, since it commutes with every element. For the case of non-degenerate particles, this can not be done due to the non-commuting nature of the non-degenerate mass matrix and the interaction matrix with off-diagonal terms. We start by defining the Heat-Kernel to be,
\begin{eqnarray}
    K(t,x,x,\Delta)=e^{-M^2t}K'(t,x,x,\Delta)\;.
\end{eqnarray}
The form of $K'(t,x,x,\Delta)$ can be obtained from the heat equation $(\partial/\partial t + \Delta )K(t,x,x,\Delta)=0$ as,
\begin{equation}\label{eq:heat_k}
    \frac{\partial}{\partial t} K'(t,x,x,\Delta) + (D^2+e^{M^2 t} U e^{-M^2t})K'(t,x,x,\Delta)=0\;.
\end{equation}
This equation marks the need for different methods of computing the HK for degenerate and non-degenerate cases. For the degenerate case, since $M$ and $U$ matrices commute, $e^{M^2 t} U e^{-M^2t} = U$, one can write $K'(t,x,x,\Delta)=b_n\,t^n/n!$, a power series in $t$ around $t=0$. This gives rise to the De-Witt recursion relation on the HKCs ($b_n$). Since in the non-degenerate case, $e^{M^2 t} U e^{-M^2t}$ matrix has an exponential in the $t$ parameter, a well-defined recursion relation between HKCs can not be obtained. 

Let us now solve Eq.~\eqref{eq:heat_k} to obtain $K'(t,x,x,\Delta)$ for a generic set of non-degenerate scalars. The differential Eq.~\eqref{eq:heat_k} has the solution,
\begin{equation}\label{eq:heat_k'}
    K'(t,x,x,\Delta)=\mathcal{T}\,\text{exp}\left[-\int_0^t  \left(D^2+e^{M^2 t'} U e^{-M^2t'}\right) dt' \right],
\end{equation}
where $\mathcal T$ represents the chronological ordering in the HK parameter $t$. $K'(t,x,x,\Delta)$ in its current form, given in  Eq.~\eqref{eq:heat_k'}, can not be expanded perturbatively.  
Heat-Kernel is defined, within the effective action,  as a spectral function of the second order differential operator acting on the fluctuations of the field, i.e., the particles in the loop. The  momentum of the particle in the loop is integrated from $-\infty$ to $\infty$. To bring  $K'(t,x,x,\Delta)$ operator in the necessary form,  we write the Heat-Kernel in the Fourier space as
\begin{align}\label{eq:path_order}
    \text{tr}\,K(t,x,x,\Delta) &=\tr \int \cfrac{d^4p}{(2\pi)^4} \bra{x} e^{-M^2t}\,\mathcal T\, \text{exp}\left[-\int_0^t \left( D^2+e^{M^2 t'} U e^{-M^2t'}\right) dt' \right] \ket{p}\bra{p}\ket{x}\;,\nn\\
    &= \text{tr}\int  \cfrac{d^4p}{(2\pi)^4}\ e^{-M^2t}\, \bra{x}\mathcal T\, \text{exp}\left[-\int_0^t  \left(D^2+e^{M^2 t'} U e^{-M^2t'}\right) dt' \right]\ket{p}\,e^{ipx}\;,\nn\\
    &= \text{tr}\int  \cfrac{d^4p}{(2\pi)^4}\ e^{-M^2t}\,  \bra{x}\ket{p}\,e^{ipx}\,\mathcal T\, \text{exp}\left[-\int_0^t \left( [D_\mu+i\,p_\mu]^2+e^{M^2 t'} U e^{-M^2t'}\right) dt' \right]\;,\nn\\
    &= \text{tr}\int  \cfrac{d^4p}{(2\pi)^4}\ e^{-M^2t}\, \,e^{p^2t}\,\mathcal T\, \text{exp}\left[-\int_0^t  \left(D^2+2i\,p\cdot D +e^{M^2 t'} U e^{-M^2t'}\right) dt' \right]\;.
\end{align}
We  redefine $p$ to absorb the $t$ parameter as $p^2t \to p^2$ in Eq.~\eqref{eq:path_order},
\begin{align}\label{eq:path_order_II}
    \text{tr}\,K(t,x,x,\Delta) = \text{tr}\int  \cfrac{d^4p}{(2\pi)^4 t^2}\ e^{-M^2t}\, e^{p^2}\,\mathcal T\, \text{exp}\left[-\int_0^t  \left(D^2+2i\,p\cdot D/\sqrt t +e^{M^2 t'} U e^{-M^2t'}\right) dt' \right],
\end{align}
and that generates Gaussian term $e^{p^2} = e^{-(p_1^2+p_2^2+p_3^2+p_4^2)}$. Note that we work here with the mostly negative Euclideanised metric, $(-,-,-,-)$. This makes the momentum integral, henceforth called the $p$-integral, convergent. Now, we define 
\begin{eqnarray}\label{eq: A_mat}
    \mathcal F(t, \mathcal A)=\mathcal T\ \text{exp}\left[-\int_0^t \mathcal A(t')\, dt'\right],
\end{eqnarray}
where
\begin{eqnarray}\label{eq:A_mat}
    \mathcal A(t) = e^{M^2 t} (D^2 + 2i\,p\cdot D/\sqrt t + U) e^{-M^2t}\;.
\end{eqnarray}
Here, $\mathcal F(t, \mathcal A)$ can be written in the form of the Volterra integral equation of the second kind as,
\begin{equation}\label{eq:volterra_eq}
    \mathcal F(t, \mathcal A) = \mathcal F(t, \mathcal A) - \int_{0}^t\,ds\,\mathcal A(s)\mathcal F(t, \mathcal A)\;,
\end{equation}
with the initial condition $\mathcal F(0, \mathcal A)=1$. We can further rewrite $\mathcal F(t, \mathcal A)$ in a series expansion  as
\begin{eqnarray}\label{eq:volterra_series}
    \mathcal F(t, \mathcal A) = 1 + \sum_{n=1}^\infty(-1)^n f_n(t,\mathcal A)\;,
\end{eqnarray}
where,
\begin{equation}\label{eq:fn}
    f_n(t,\mathcal A)=\int_{0}^{t}ds_1\int_{0}^{s_1}ds_2\cdots\int_{0}^{s_{n-1}}ds_n \,\mathcal A(s_1)\mathcal A(s_2)\cdots \mathcal A(s_n)\;.
\end{equation}
Using the derived HK  in Eqs.~\eqref{eq:path_order}-\eqref{eq:fn}, the one-loop effective Lagrangian, see Eq.~\eqref{eq:effective-action-with-HK}, takes the following form
\begin{eqnarray}\label{eq:final_exp}
     \L_{\text{eff}} &=& c_s\, \text{tr}\int_0^\infty \frac{dt}{t}\int  \frac{d^4p}{(2\pi)^4\, t^2}\ e^{p^2}\,e^{-M^2t}\,\left[1 + \sum_{n=1}^\infty(-1)^n f_n(t,\mathcal A)\right]\;.
\end{eqnarray}
It is evident that setting $ \mathcal F(t, \mathcal A)=1$, the $p$ and $t$-integrals lead to the free particle Heat-Kernel $K_0(t,x,x,\partial^2+M^2)$ in the non-degenerate case. Here, the key distinguishing feature is that the non-commuting nature of the interaction matrix and the non-degenerate mass matrix introduces the $t$ parameter dependence in the series expansion ($f_n$) in a non-trivial way compared to the degenerate case ($b_n\equiv \text{HKC}$) \cite{Banerjee:2023iiv,Chakrabortty:2023yke}. This is also the reason that we cannot separate out the free and interaction parts unlike the degenerate cases \cite{Banerjee:2023iiv,Chakrabortty:2023yke}. It is worth noting that in the degenerate limit,
\begin{eqnarray}
    f_n(t,\mathcal A) \longrightarrow \frac{t^n}{n!}\ b_n\;.
\end{eqnarray}

%%%%%%%%%%%%%%%%%%%%%%%%%%%%%%%%%%%%%%%%%%%%%%%%
\section{Methodology and Results}
\label{sec:procedure}
%%%%%%%%%%%%%%%%%%%%%%%%%%%%%%%%%%%%%%%%%%%%%%%%%%%%%

Before going into the details of obtaining the effective action operators, let us discuss the notations and terminologies used. We categorize the one-loop effective action operators based on the number of derivatives and the interaction terms they contain. An operator containing, for instance, $r$ covariant derivatives ($D_\mu$) and $s$ interaction terms ($U_{ij}$) belongs to the class $\mathcal D^rU^s$. The mass dimension of the operators of an operator class is not very straightforward. It depends on the mass dimension of coupling in the interaction term $U_{ij}$. For the dimension counting of an operator, we consider contribution only from fields and covariant derivatives, and not from the mass dimension of coupling constants. Hence, $U_{ij}$ with dimensionless coupling contributes a mass dimension of two towards the operator dimension counting while a $U_{ij}$ with a coupling constant of dimension one, tri-linear interactions, contributes a dimension of one. When referring to operators up to a particular dimension, we include operators with $U_{ij}$ contributing dimension one to the operator mass dimension.

Here, we point out the   key aspects of our methodology:
\begin{itemize}
    \item It is important, first, to identify the orders of $f_n(t,\mathcal A)$, in Eq.~\eqref{eq:fn}, to encapsulate all the contributions to construct operators of a specific mass dimension.
    \item Then, after obtaining the open derivative structures of the operators belonging to a specific class, we need to translate them into their possible closed derivative forms.
    \item It is advised to compute the operators for the complete non-degenerate case as it provides the result in  most generic form.
    \item Depending on the nature mass-degeneracy of the UV-theory, we can obtain the results from the generic one by choosing suitable degenerate-mass limits.
    \item Heavy-Light mixing effects can be captured following the similar procedure by setting light mass $\to 0$ after assuring the infrared safety of the theory.
\end{itemize}
These points are equally applicable for scalar and fermion fields. We elaborate these points in detail in the following subsections.
\subsection{Overview on obtaining the effective action operators}

To identify the order of $f_n(t,\mathcal A)$ up to which one has to compute in Eq.~\eqref{eq:fn}, let us first focus on the detailed structure of the matrix $\mathcal A$. The elements of the matrix $\mathcal A$ have three different kinds of sub-structures, $D^2$, $p\cdot D$, and $U_{ij}$. Hence, $\mathcal A^n$ can at most contribute $D^{2n}$, $D^n$ or $U^n_{ij}$ to an operator. Since operators of class $\mathcal D^rU^s$ gets contribution from all powers of $\mathcal A$ ranging from $\mathcal A^{r/2+s}$ to $\mathcal A^{r+s}$, it is necessary to compute all orders of $f_n(t,\mathcal A)$ from $f_{r/2+s}(t,\mathcal A)$ to $f_{r+s}(t,\mathcal A)$ to capture the complete contribution to the operator class.

 After performing the $p$-integral and $t$-integral, the operators obtained from the methodology outlined above, are given in their open derivative forms. To reduce them into the covariant structures, where $D_{\mu}$'s appear only in commutators, one can use the covariant operators obtained in Ref.~\cite{Banerjee:2023iiv}, with unknown coefficients and proper particle indices ($i,j$) on $U$, as an operator basis and solve for the coefficients. This can be done by expanding the covariant structures of the basis into open derivative structures and comparing their respective coefficients with those obtained from the program. This is similar to the procedure used in Refs.~\cite{Banerjee:2023iiv,Zhang:2016pja} for obtaining the closed derivative structure when using the covariant diagram approach.

It is important to note that in the non-degenerate case, the operator basis is obtained by introducing the particle indices ($i$, $j$) on $U$, from Ref.~\cite{Banerjee:2023iiv}, is not an independent minimal basis. Hence, when solving for the unknown coefficients the number of linear equations required might be less than the unknowns. In such cases, one has to identify the independent operators from among the basis obtained from Ref.~\cite{Banerjee:2023iiv}. The fact that hermitian conjugates have the same coefficients since they are real, might help reduce the number of unknowns in some cases.

\subsection{Effective Operators with two non-degenerate Scalars: an example}
\label{sec:2_paticle_result}
Below we provide the UOLEA operators for the case of two non-degenerate particles up to mass dimension six. The Lagrangian for this case is given by,
\begin{eqnarray}
    \L = \Phi_1^\dagger (D^2 +M_1^2 +U_{11})\Phi_1 + \Phi_2^\dagger (D^2 +M_2^2 +U_{22})\Phi_2 + \Phi_1^\dagger U_{12}\Phi_2 + \Phi_2^\dagger U_{21}\Phi_2\;.
\end{eqnarray}
The $\mathcal A$ matrix in Eq.~\eqref{eq:A_mat} is given by,
\begin{eqnarray}
    \mathcal A = \begin{bmatrix}
        D^2 + 2i p\cdot D/\sqrt{t} + U_{11}&U_{12}\ e^{\Delta_{12}^2t}\\
        U_{21}\ e^{-\Delta_{12}^2 t}&D^2 + 2i p\cdot D/\sqrt{t} + U_{22}
    \end{bmatrix}.
\end{eqnarray}
The Wilson coefficients (WCs) of the UOLEA operators with the pure heavy-loops, with only one kind of particle in the loops, are the same as the results provided in Ref.~\cite{Banerjee:2023iiv}. Here we provide only the results for the mixed loop UOLEA and we do not present the results separately for operators that are different under the trace but can be obtained just by interchanging indices.

For the purpose of the presentation, let us define the mass-squared separation as $\Delta^2_{ij}=M_i^2-M_j^2$. In what follows we first discuss from which order of $f_n(t,\mathcal{A})$ each of these classes can arise and then present their corresponding covariant structures and Wilson coefficients.

\subsection*{\underline{$\mathcal{O}(U^2)$}}
As discussed above, this class of operators gets contribution from only $f_2(t,\mathcal A)$.

\begin{itemize}
\item \underline{Covariant operators:}
\begin{eqnarray}
    \L_{\eff}[U^2] = \C_{ij}\,\text{tr}(U_{ij}U_{ji})\, .
\end{eqnarray}
\item \underline{Wilson coefficients:}\footnote{We do not mention the factor of $\frac{cs}{(4\pi)^2}$ in the Wilson coefficients explicitly.}
\begin{align}
    &\C_{12} = 1-\frac{M_1^2\log{\cfrac{M_1^2}{\mu^2}}}{\Delta_{12}^2}+\frac{M_2^2\log{\cfrac{M_2^2}{\mu^2}}}{\Delta_{12}^2}\, .\nn
\end{align}
\end{itemize}
%%%%%%%%%%%%%%%%%%%%%%%%%%%%%%%%%%%%%%%%%%%%%%%%%%%%%%%%%%%%%%%%%
\subsection*{\underline{$\mathcal{O}(U^3)$}}
This class of operators gets contribution from only $f_3(t,\mathcal A)$.

\begin{itemize}
\item \underline{Covariant operators:}
\begin{eqnarray}
    \L_{\eff}[U^3] = \C_{ijk}\,\text{tr}(U_{ij}U_{jk}U_{ki})\, .
\end{eqnarray}
\item \underline{Wilson coefficients:}
\begin{align}
    &\C_{122} = \frac{1}{\Delta^2_{12}}-\frac{M_1^2\log{\frac{M_1^2}{M^2_2}}}{(\Delta_{12}^2)^2}\, .\nn
\end{align}
\end{itemize}
%%%%%%%%%%%%%%%%%%%%%%%%%%%%%%%%%%%%%%%%%%%%%%%%%%%%%%%%%%%%%%%%%
\subsection*{\underline{$\mathcal{O}(U^4)$}}
This class of operators gets contribution from only $f_4(t,\mathcal A)$.

\begin{itemize}
    \item \underline{Covariant operators:}
\begin{align}
    \L_\eff[U^4] =& \C_{ijkl}\ \tr(U_{ij}U_{jk}U_{kl}U_{li})\, .
\end{align}
   \item \underline{Wilson coefficients:}
\begin{gather}\small
    \C_{1122} = -\frac{2}{(\Delta^2_{12})^2}+\frac{(M_1^2+M_2^2)\log{\frac{M_1^2}{M^2_2}}}{(\Delta^2_{12})^3}\, ,\quad\C_{1212} = -\frac{1}{(\Delta^2_{12})^2}+\frac{(M_1^2+M_2^2)\log{\frac{M_1^2}{M^2_2}}}{2(\Delta^2_{12})^3}\,,\nn\\
    \C_{1112}=\frac{M_1^2+M_2^2}{2M_1^2(\Delta^2_{12})^2}-\frac{M_2^2\log{\frac{M_1^2}{M^2_2}}}{(\Delta^2_{12})^3}\, .\nn
\end{gather}
\end{itemize}
%%%%%%%%%%%%%%%%%%%%%%%%%%%%%%%%%%%%%%%%%%%%%%%%%%%%%%%%%%%%%%%%%
\subsection*{\underline{$\mathcal{O}(U^5)$}}
This class of operators gets contribution from only $f_5(t,\mathcal A)$.

\begin{itemize}
    \item \underline{Covariant operators:}
\begin{eqnarray}
    \L_\eff[U^5] =& \C_{ijklm}\tr(U_{ij}U_{jk}U_{kl}U_{lm}U_{mi})\, .
\end{eqnarray}
  \item  \underline{Wilson coefficients:}
\begin{eqnarray}
    \C_{12222} &=& -\frac{M_1^4-5M_1^2M_2^2-2M_2^4}{6 M_2^4(\Delta^2_{12})^3}-\frac{M_1^2\log{\frac{M_1^2}{M^2_2}}}{(\Delta^2_{12})^4}\, ,\nn\\
    \quad\C_{11122} &=& \frac{5M_1^2+M_2^2}{2 M_1^2(\Delta^2_{12})^3}-\frac{(M_1^2+2M_2^2)\log{\frac{M_1^2}{M^2_2}}}{(\Delta^2_{12})^4}\, ,\nn\\
    \C_{11212} &=& \frac{5M_1^2+M_2^2}{2 M_1^2(\Delta^2_{12})^3}-\frac{(M_1^2+2M_2^2)\log{\frac{M_1^2}{M^2_2}}}{(\Delta^2_{12})^4}\,.\nn
\end{eqnarray}
\end{itemize}
%%%%%%%%%%%%%%%%%%%%%%%%%%%%%%%%%%%%%%%%%%%%%%%%%%%%%%%%%%%%%%%%%
\subsection*{\underline{$\mathcal{O}(U^6)$}}
This class of operators gets contribution from only $f_6(t,\mathcal A)$

\begin{itemize}
    \item \underline{Covariant operators:}
\begin{align}
    \L_\eff[U^6] =&\, \C_{ijklmn}\tr(U_{ij}U_{jk}U_{kl}U_{lm}U_{mn}U_{ni})\, .
\end{align}
  \item \underline{Wilson coefficients:}
\begin{align}
    &\C_{111122}=\C_{111212}=2\ \C_{112112} = \frac{-17M_1^4-8M_1^2M_2^2+M_2^4}{6 M_1^4(\Delta^2_{12})^4}+\frac{(M_1^2+3M_2^2)\log{\frac{M_1^2}{M^2_2}}}{(\Delta^2_{12})^5}\, ,\nn\\
    &\C_{112122}= \C_{111222}= 3\ \C_{212121}=\frac{M_1^4+10M_1^2M_2^2+M_2^4}{2 M_1^2M_2^2(\Delta^2_{12})^4}-\frac{3(M_1^2+M_2^2)\log{\frac{M_1^2}{M^2_2}}}{(\Delta^2_{12})^5}\, .\nn
\end{align}
\end{itemize}
%%%%%%%%%%%%%%%%%%%%%%%%%%%%%%%%%%%%%%%%%%%%%%%%%%%%%%%%%%%%%%%%%
\subsection*{\underline{$\mathcal{O}(\mathcal{D}^2U^2)$}}
These operators get contribution from $f_3(t,\mathcal A)$ and $f_4(t,\mathcal A)$.

\begin{itemize}
    \item \underline{Covariant operators:}
\begin{eqnarray}
    \mathcal{L}_{\text{eff}}[\mathcal{D}^2U^2] = \C_{ij}\,\text{tr}([D_{\mu},U_{ij}][D_{\mu},U_{jk}])\, .
\end{eqnarray}
\item \underline{Wilson coefficients:}
\begin{align}
    &\C_{12} = \cfrac{M_1^2+M_2^2}{2(\Delta_{12}^2)^2}-\cfrac{M_1^2M_2^2\log\frac{M_1^2}{M_2^2}}{(\Delta_{12}^2)^3}\, .\nn
\end{align}
\end{itemize}
%%%%%%%%%%%%%%%%%%%%%%%%%%%%%%%%%%%%%%%%%%%%%%%%%%%%%%%%%%%%%%%%%
\subsection*{\underline{$\mathcal{O}(\mathcal{D}^2U^3)$}}
These operators get contribution from $f_4(t,\mathcal A)$ and $f_5(t,\mathcal A)$.

\begin{itemize}
    \item \underline{Covariant operators:}
\begin{eqnarray}
    \mathcal{L}_{\text{eff}}[\mathcal{D}^2U^3] =& \C_{ijk} \tr(U_{ij}[D_\mu,U_{jk}][D_\mu,U_{ki}])\, .
\end{eqnarray}
    \item \underline{Wilson coefficients:}
\begin{align}
    &\C_{112} = -\frac{M_1^2+5 M_2^2}{2 \left(\Delta_{12}^2\right)^3}+\frac{M_2^2 \left(2 M_1^2+M_2^2\right) \log \left(\frac{M_1^2}{M_2^2}\right)}{\left(\Delta_{12}^2\right)^4}\, ,\nn\\
    &\C_{211}=\C_{121} =- \frac{2 M_1^4+5 M_1^2 M_2^2-M_2^4}{6 M_1^2 \left(\Delta_{12}^2\right)^3}+\frac{M_1^2 M_2^2 \log \left(\frac{M_1^2}{M_2^2}\right)}{\left(\Delta_{12}^2\right)^4}\, .\nn
\end{align}
\end{itemize}
%%%%%%%%%%%%%%%%%%%%%%%%%%%%%%%%%%%%%%%%%%%%%%%%%%%%%%%%%%%%%%%%%
\subsection*{\underline{$\mathcal{O}(\mathcal{D}^2U^4)$}}
These operators get contribution from $f_5(t,\mathcal A)$ and $f_6(t,\mathcal A)$.
\begin{itemize}
\item \underline{Covariant operators:}
\begin{eqnarray}
    \mathcal{L}_{\text{eff}}[\mathcal{D}^2U^4] = \C^{(1)}_{ijkl}\ U_{ij}U_{jk}[D_\mu,U_{kl}][D_\mu,U_{li}] + \C^{(2)}_{ijkl}\ U_{ij}[D_\mu,U_{jk}]U_{kl}[D_\mu,U_{li}] \, .
\end{eqnarray}
\item \underline{Wilson coefficients:}
\begin{align}
    &\C^{(1)}_{1112} = \frac{M_1^4+10 M_1^2 M_2^2+M_2^4}{2 M_1^2 \left(\Delta^2_{12}\right)^4}-\frac{3 M_2^2 \left(M_1^2+M_2^2\right) \log\frac{M_1^2}{M_2^2}}{ \left(\Delta^2_{12}\right)^5}\, ,\nn\\
    &\C^{(1)}_{1221} = -\frac{17 M_1^4+8 M_1^2 M_2^2-M_2^4}{6 M_1^2 \left(\Delta^2_{12}\right)^4}+\frac{ M_1^2 \left(M_1^2+3 M_2^2\right) \log\frac{M_1^2}{M_2^2}}{ \left(\Delta^2_{12}\right)^5}\, , \nn\\
    &\C^{(1)}_{1222} = \frac{M_1^6-5 M_1^4 M_2^2+13 M_1^2 M_2^4+3 M_2^6}{12 M_2^4 \left(\Delta^2_{12}\right)^4}-\frac{ M_1^2 M_2^2 \log\frac{M_1^2}{M_2^2}}{ \left(\Delta^2_{12}\right)^5}\, ,\nn\\
    &\C^{(2)}_{1122}= -\frac{6 \left(M_1^2+M_2^2\right)}{\left(\Delta^2_{12}\right)^4}+\frac{2 \left(M_1^4+4 M_1^2 M_2^2+M_2^4\right) \log\frac{M_1^2}{M_2^2}}{\left(\Delta^2_{12}\right)^5}\, ,\nn\\
    &\C^{(1)}_{1121} = \C^{(1)}_{1211} = \frac{3 M_1^6+13 M_1^4 M_2^2-5 M_1^2 M_2^4+M_2^6}{12M_1^4 \left(\Delta^2_{12}\right)^4}-\frac{M_1^2 M_2^2 \log\frac{M_1^2}{M_2^2}}{\left(\Delta^2_{12}\right)^5}\, ,\nn\\
    &\C^{(1)}_{1122} = \C^{(1)}_{1212} = -\frac{-M_1^4+8 M_1^2 M_2^2+17 M_2^4}{6 M_2^2 \left(\Delta^2_{12}\right)^4} + \frac{ M_2^2 \left(3 M_1^2+M_2^2\right) \log\frac{M_1^2}{M_2^2}}{ \left(\Delta^2_{12}\right)^5}\, , \nn\\
    & \C^{(2)}_{1112} = \C^{(2)}_{2111}=\frac{5 M_1^6+33 M_1^4 M_2^2-3 M_1^2 M_2^4+M_2^6}{12 M_1^4\left(\Delta^2_{12}\right)^4}-\frac{M_2^2 \left(2 M_1^2+M_2^2\right) \log\frac{M_1^2}{M_2^2}}{\left(\Delta^2_{12}\right)^5}\, ,\nn\\
    & \C^{(2)}_{1212}=\C^{(2)}_{1221}=\frac{M_1^6-7 M_1^4 M_2^2-7 M_1^2 M_2^4+M_2^6}{6 M_1^2 M_2^2 \left(\Delta^2_{12}\right)^4}+\frac{2 M_1^2 M_2^2  \log\frac{M_1^2}{M_2^2}}{\left(\Delta^2_{12}\right)^5}\, .\nn\\
\end{align}
\end{itemize}
%%%%%%%%%%%%%%%%%%%%%%%%%%%%%%%%%%%%%%%%%%%%%%%%%%
\subsection*{\underline{$\mathcal{O}(\mathcal{D}^2U^5)$}}
These operators get contribution from $f_6(t,\mathcal A)$ and $f_7(t,\mathcal A)$.
\begin{itemize}
\item \underline{Covariant operators:}
\begin{align}
    \mathcal{L}_{\text{eff}}[\mathcal{D}^2U^5] =&\, \C^{(1)}_{ijklm}\,\tr(U_{ij}U_{jk}U_{kl}U_{lm}\,[D_{\mu},[D_{\mu},U_{mi}]])\nn\\
    &+ \;\C^{(2)}_{ijklm} \,\tr(U_{ij}U_{jk}U_{kl}[D_{\mu},U_{lm}][D_{\mu},U_{mi}])\}\, .
\end{align}
\item \underline{Wilson coefficients:}
\begin{align}
    &\C^{(1)}_{12121}=-\frac{77 M_1^6+105 M_1^4 M_2^2-3 M_1^2 M_2^4+M_2^6}{24 M_1^4 (\Delta^2_{12})^5}+\frac{\left(2 M_1^4+10 M_1^2 M_2^2+3 M_2^4\right) \log \frac{M_1^2}{M_2^2}}{2 (\Delta^2_{12})^6}\, ,\nn\\
    &\C^{(1)}_{12211}=-\frac{17 M_1^4+8 M_1^2 M_2^2-M_2^4}{12 M_1^4 (\Delta^2_{12})^4}+\frac{\left(M_1^2+3 M_2^2\right) \log \frac{M_1^2}{M_2^2}}{2 (\Delta^2_{12})^5}\, ,\nn\\
    &\C^{(2)}_{12121}=-\frac{10 M_1^4+19 M_1^2 M_2^2+M_2^4}{3 M_1^2 (\Delta^2_{12})^5}+\frac{\left(M_1^4+6 M_1^2 M_2^2+3 M_2^4\right) \log \frac{M_1^2}{M_2^2}}{(\Delta^2_{12})^6}\, ,\nn\\
    &\C^{(2)}_{12211}= -\frac{M_2^2 \left(2 M_1^2+3 M_2^2\right) \log \frac{M_1^2}{M_2^2}}{(\Delta^2_{12})^6}+\frac{3 M_1^6+47 M_1^4 M_2^2+11 M_1^2 M_2^4-M_2^6}{12 M_1^4 (\Delta^2_{12})^5}\, .\nn
\end{align}
\end{itemize}
%%%%%%%%%%%%%%%%%%%%%%%%%%%%%%%%%%%%%%%%%%%%%%%%%%%%%%%%%%%%%%%%%
\subsection*{\underline{$\mathcal{O}(\mathcal{D}^4U^2)$}}
These operators get contribution from $f_4(t,\mathcal A)$, $f_5(t,\mathcal A)$ and $f_6(t,\mathcal A)$.
\begin{itemize}
\item \underline{Covariant operators:}
\begin{align}
    \mathcal{L}_{\text{eff}}[\mathcal{D}^4U^2] =&\, \C^{(1)}_{ij}\,\tr([D_{\mu},[D_{\mu},U_{ij}]][D_{\nu},[D_{\nu},U_{ji}]])+ \C^{(2)}_{ij} \text{tr}([D_{\mu},U_{ij}][D_{\nu},U_{ji}]G_{\nu\mu}) \nn\\
    & + \C^{(3)}_{ij} \text{tr}(U_{ij}U_{ji}G_{\nu\mu}G_{\nu\mu})+\C^{(4)}_{ij} \tr\{(U_{ij}[D_{\mu},U_{ji}]-[D_{\mu},U_{ij}]U_{ji})[D_{\nu},G_{\nu\mu}]\}\, .
\end{align}
\item \underline{Wilson coefficients:}
\begin{align}
    &\C^{(1)}_{12} =\C^{(2)}_{12} = \frac{M_1^4+10M_1^2M_2^2+M_2^2}{6(\Delta^2_{12})^4} -\frac{M_1^2M_2^2(M_1^2+M_2^2)\log\frac{M_1^2}{M_2^2}}{(\Delta^2_{12})^5}\, ,\nn\\
    &\C^{(3)}_{12} = \frac{2M_1^4+5M_1^2M_2^2-M_2^2}{12M_1^2(\Delta^2_{12})^3} -\frac{M_1^2M_2^2\log\frac{M_1^2}{M_2^2}}{2(\Delta^2_{12})^4}\, ,\nn\\
    &\C^{(4)}_{12} = \frac{-2M_1^4-11M_1^2M_2^2+7M2^2}{18(\Delta^2_{12})^4} -\frac{M_2^2(-3M_1^4+M_2^4)\log\frac{M_1^2}{M_2^2}}{6(\Delta^2_{12})^4}\, .\nn
\end{align}
\end{itemize}
%%%%%%%%%%%%%%%%%%%%%%%%%%%%%%%%%%%%%%%%%%%%%%%%%%%%%%%%%%%%%%%%%
\subsection*{\underline{$\mathcal{O}(\mathcal{D}^4U^3)$}}
These operators get contribution from $f_5(t,\mathcal A)$, $f_6(t,\mathcal A)$ and $f_7(t,\mathcal A)$.
\begin{itemize}
\item \underline{Covariant operators:}
\begin{align}
    \mathcal{L}_{\text{eff}}[\mathcal{D}^4U^3] =&\, \C^{(1)}_{ijk}\,\tr(U_{ij}U_{jk}U_{ki}G_{\mu\nu}G_{\mu\nu})+\C^{(2)}_{ijk}\,\tr(U_{ij}[D_\mu,[D_\mu,U_{jk}]][D_\nu,[D_\nu,U_{ki}]])\nn\\
    &+\C^{(3)}_{ijk}\,\tr(U_{ij}G_{\mu\nu}[D_\mu,U_{jk}][D_\nu,U_{ki}]+U_{ji}[D_\mu,U_{ik}][D_\nu,U_{kj}]G_{\mu\nu})\nn\\
    &+\C^{(4)}_{ijk}\,\tr([D_\mu,[D_\mu,U_{ij}]][D_\nu,U_{jk}][D_\nu,U_{ki}])+\C^{(5)}_{ijk}\,\tr(U_{ij}U_{jk}G_{\mu\nu}U_{ki}G_{\mu\nu})\nn\\
    &+\C^{(6)}_{ijk}\,\tr(U_{ij}U_{jk}[D_\nu,U_{ki}][D_\mu,G_{\mu\nu}]-U_{kj}U_{ji}[D_\mu,G_{\mu\nu}][D_\nu,U_{ik}]).
\end{align}
\item \underline{Wilson coefficients:} We provide the results for only a subset of operators of this class.
\begin{align}
    &\C^{(1)}_{121}= -\frac{M_1^2 M_2^4 \log \left(\frac{M_1^2}{M_2^2}\right)}{(\Delta^2_{12})^6}-\frac{3 M_1^8-27 M_1^6 M_2^2-47 M_1^4 M_2^4+13 M_1^2 M_2^6-2 M_2^8}{60 M_1^4 (\Delta^2_{12})^5}\, ,\nn\\
    &\C^{(2)}_{121}= -\frac{9 M_1^8+104 M_1^6 M_2^2+4 M_1^4 M_2^4+4 M_1^2 M_2^6-M_2^8}{60 M_1^4 (\Delta^2_{12})^5}+\frac{M_1^2 M_2^2 \left(M_1^2+M_2^2\right) \log \left(\frac{M_1^2}{M_2^2}\right)}{(\Delta^2_{12})^6}\, ,\nn\\
    &\C^{(3)}_{121}= -\frac{M_1^4+19 M_1^2 M_2^2+10 M_2^4}{9 (\Delta^2_{12})^5}+\frac{M_2^2 \left(3 M_1^4+6 M_1^2 M_2^2+M_2^4\right) \log \left(\frac{M_1^2}{M_2^2}\right)}{3 (\Delta^2_{12})^6}\, ,\nn\\
    &\C^{(4)}_{121}= -\frac{2(M_1^4+19 M_1^2 M_2^2+10 M_2^4)}{9 (\Delta^2_{12})^5}+\frac{2M_2^2 \left(3 M_1^4+6 M_1^2 M_2^2+M_2^4\right) \log \left(\frac{M_1^2}{M_2^2}\right)}{3 (\Delta^2_{12})^6}\, .\nn
\end{align}
\end{itemize}
%%%%%%%%%%%%%%%%%%%%%%%%%%%%%%%%%%%%%%%%%%%%%%%%%%
\subsection{Paving the path from non-Degenerate to Degenerate case}
In case of the spectrum of the theory possess some degeneracy, then one can employ suitable limits on masses, in the non-degenerate result, to obtain the effective action for that theory. It is important to recall that the most generic result can be computed when the number of non-degenerate particles equals to the number of interaction term $U_{ij}$ in the operator of interest. For example, if we are interested in the operators of class $\mathcal D^r U^s$ then we, first, obtain the result with $s$ non-degenerate particles. Then, for any other theory having non-degeneracy of order $\alpha < s$, can be computed from the earlier one employing suitable limits. 

Here, we demonstrate that with an example case. The two-particle results provided in Sec.~\ref{sec:2_paticle_result} can be obtained from a more compact general result of $N$ particles which have been presented in Appendix \ref{app:B}. To demonstrate taking the degenerate limit from the general results, let us take the case of $\mathcal{D}^2U^3$ operator class. The complete non-degenerate case result for the operator in class $\mathcal D^2U^3$ is given by,
\begin{align}\label{eq:n_general}
    \C_{ijk}=&-\frac{M_i^2 \left(M_j^2+M_k^2\right)+M_k^2 \left(M_j^2-3 M_k^2\right)}{2 (\Delta_{ik}^2)^2 (\Delta_{jk}^2)^2} +\frac{ M_i^2 M_k^2 \log M_i^2}{\Delta_{ij}^2(\Delta_{ki}^2)^3}+\frac{ M_j^2 M_k^2 \log M_j^2}{\Delta_{ij}^2(\Delta_{jk}^2)^3}\nn\\&- \frac{M_k^2 \left(M_i^2 M_j^2 \left(M_i^2+M_j^2-3 M_k^2\right)+M_k^6\right)\log M_k^2}{(\Delta_{ik}^2)^3 (\Delta_{jk}^2)^3}\, .
\end{align}
Here, the $i,j,k$ represent the particle indices and each of them runs over all the particles in the Lagrangian. The coefficients $\C_{112}$, $\C_{211}$ and $\C_{121}$, in the two-particle scenario, can be obtained from the operator coefficient in Eq.~\eqref{eq:n_general}, by taking the limit ($i = j = 1$, $k=2$), ($j = k = 1$, $i=2$) and ($i = k = 1$, $j=2$) respectively. For the case of more than three particles, the coefficients will just differ by the particle index, the coefficient structure would not change. The two-particle results provided in the previous section can be verified by taking the degenerate limit in the results provided in Ref. \cite{Zhang:2016pja} \footnote{Note the difference in notation used here and Ref.~\cite{Zhang:2016pja}. They use $P_\mu=iD_\mu$ as the covariant derivative operator.}. 
%%%%%%%%%%%%%%%%%%%%%%%%%%%%%%%%%%%%%%%%%%%%%%%%%%
\subsection{Light-Heavy Mixing Effects: a special case of non-Degenerate to Degenerate}
\label{subsec:Heavy_Light}
%%%%%%%%%%%%%%%%%%%%%%%%%%%%%%%%%%%%%%%%%%%%%%%%%%
The case of integrating out fluctuations of light particles along with heavy particles, i.e. the heavy-light mixed loops, can be looked at as a special case of the non-degenerate effective action. In writing Eq.~\eqref{eq:final_exp}, it assumes that $M$ is the largest scale and hence includes the loop momentum up to $M$. The $e^{-M^2t}$ acts as a regulator and the higher dimension operators are scaled as $1/M^n$. This is no longer the case when considering light particles in the loop. In the limit light particle mass is zero, additional IR divergences emerge. In Ref.~\cite{Zhang:2016pja}, the heavy-light mixing contribution is derived as the hard momentum limit expansion of the light particle propagator, which includes momentum greater than the mass of the light particle. This is in contradiction if we just use $e^{-M^2t}$ as the regulator. So in order to accommodate for light propagator momentum in the hard limit, we add a Gaussian IR-regulator $e^{-\Lambda^2 t}$, to the $t$ integral in Eq.~\eqref{eq:final_exp}, which is usually of the order of the mass of the lightest heavy particle in the Lagrangian,
\begin{eqnarray}\label{eq:final_exp_HL}
     \L_{\text{eff}} &=& c_s\, \text{tr}\int_0^\infty \frac{dt}{t}\int  \frac{d^4p}{(2\pi)^4\, t^2}\ e^{p^2}\,e^{-(\Lambda^2+M^2)t}\,\left[1 + \sum_{n=1}^\infty(-1)^n f_n(t,\mathcal A)\right]\;.
\end{eqnarray}
In the limit that the light particle mass is zero, the addition of this regulator aids in including the hard momentum region for the light propagator. For the heavy particles, we take $M_i^2+\Lambda^2\rightarrow M_i^2$ as $\Lambda$ is of the order of lightest heavy particle mass, and for the case of light particles, we take $m_i^2+\Lambda^2\rightarrow \Lambda^2$. For matching with the low energy theory, we overlook the IR divergent terms (as it is there in the IR theory as well) and consider only the finite piece.

Since this method does not use the classical solution of the heavy particle explicitly or perform any local expansion of the non-local operators, we do not have to go through the cumbersome method of calculating non-local contributions \cite{Henning:2016lyp} or subtraction terms as calculated in Ref.~\cite{Ellis:2016enq}. The subtraction terms calculated in Ref.~\cite{Ellis:2016enq} are equivalent to the extraction of external momentum-dependent terms in the amplitude and in the current method, the inclusion of an IR regulator directly gives the subtraction terms as the regulator-dependent terms. Hence, the above-described methodology can be generalised to the heavy-light mixed loops in a straightforward manner.

We denote the light particle by the index $L$ and the heavy particle by $H$. An explicit calculation of the heavy-light mixing for the operator $U_{HL}[D_\mu,U_{LL}][D_\mu,U_{LH}] $ in class $\mathcal D^2 U^3$, is given in Appendix \ref{app:A}. Here we provide the coefficients \footnote{Henceforth we do not explicitly write the UV scale ($\mu$). Terms such as $\log \frac{M^2}{\mu^2}$ are written as $\log M^2$. The UV scale dependence can be easily identified through dimensional considerations.} of the light-heavy mixed loops for a few classes of operators that have been presented in Sec. \ref{sec:2_paticle_result}. 

\subsection*{\underline{$\mathcal{O}(\mathcal{D}^2U^2)$}}
\begin{align}
    \C_{LH} = \frac{1}{2M^2}\, .
\end{align}

\subsection*{\underline{$\mathcal{O}(\mathcal{D}^2U^3)$}}
\begin{align}
    &\C_{LHH} =  -\frac{1}{3M^4}\, ,& &\C_{HLL} =\frac{1}{2M^4}\, ,&
    &\C_{LLH} = \frac{1}{M^4}\left(\frac{5}{2}-\log M^2\right)\, ,& &\C_{HHL} = -\frac{1}{2M^4}\, .& 
\end{align}

\subsection*{\underline{$\mathcal{O}(\mathcal{D}^2U^4)$}}
\begin{gather}
    \C^{(1)}_{LHLH} =\C^{(1)}_{LLHH} = \C^{(1)}_{HLLH} =-\frac{17}{6 M^6} +\frac{\log M^2}{M^6}\, ,\quad
    \C^{(1)}_{HLHL} =\C^{(1)}_{HHLL} =\C^{(1)}_{LHHL} =-\frac{{1}}{M^6}\, ,\nn\\
    \C^{(1)}_{LLLH} = \frac{6}{M^6}-\frac{3 \log M^2}{M^6}\, ,\quad
    \C^{(1)}_{HHHL} =\frac{1}{2M^6}\, ,\quad
    \C^{(1)}_{HHLH} =\C^{(1)}_{HLHH} = \frac{1}{4M^6}\, ,\nn\\
    \C^{(1)}_{LHLL} =\C^{(1)}_{HLLL} = \C^{(1)}_{LLHL} =\frac{2}{3 M^6}\, ,\quad
    \C^{(2)}_{LHLH} = \C^{(2)}_{HLHL} = -\frac{5}{12M^6}\, ,\quad  \C^{(2)}_{LHHL}=-\frac{5}{6 M^6}\, ,  \nn\\
    \C^{(2)}_{LLLH}=\C^{(2)}_{HLLL}= \frac{8}{3 M^6}-\frac{\log M^2}{M^6}\, ,\quad
    \C^{(2)}_{HHHL}=\C^{(2)}_{LHHH}= \frac{5}{12 M^6}\, ,\quad
    \C^{(2)}_{LLHH}= -\frac{6}{M^6}+\frac{2 \log M^2}{M^6}\, .
\end{gather}

\subsection*{\underline{$\mathcal{O}(\mathcal{D}^2U^5)$}}
\begin{gather}
    \C^{(1)}_{LHLHL}=-\frac{3 \log M^2-9}{2 M^8}\, ,
    \,\,\C^{(1)}_{HLHLH}= -\frac{77-24 \log M^2}{24 M^8}\, ,
    \,\,\C^{(1)}_{LHHLL}= -\frac{3-3 \log M^2}{2 M^8}\, ,\nn\\
    \,\,\C^{(1)}_{HLLHH}= -\frac{17-6 \log M^2}{12 M^8}\, ,
    \,\,\C^{(2)}_{LHLHL}= \frac{7-3 \log M^2}{M^8}\, ,
    \,\,\C^{(2)}_{HLHLH}=- \frac{10-3 \log M^2}{3 M^8}\, ,\nn\\
    \,\,\C^{(2)}_{LHHLL}=- \frac{5-3 \log M^2}{M^8}\, ,
    \,\,\C^{(2)}_{HLLHH}= \frac{1}{4 M^8}\, .\nn
\end{gather}

\subsection*{\underline{$\mathcal{O}(\mathcal{D}^4U^2)$}}
\begin{align}
    \C^{(1)}_{LH} = \frac{1}{6M^4}\, ,
    \,\,\C^{(2)}_{LH} = \frac{1}{6M^4}\, ,
    \,\,\C^{(2)}_{HL} = \frac{1}{6M^4}\, ,
    \,\,\C^{(3)}_{LH} = -\frac{1}{4 M^4}\, \nn\\
    \C^{(3)}_{HL} = \frac{1}{6M^4}\, ,\,\,\C^{(4)}_{LH} = \frac{-3 \log M^2 +7}{18 M^4}\, ,\,\,
    \C^{(4)}_{HL} = -\frac{1}{9M^4}\, .
\end{align}
\subsection*{\underline{$\mathcal{O}(\mathcal{D}^4U^3)$}}
\begin{align}
    &\C^{(1)}_{HLH}=  -\frac{1}{20 M^6}\, ,&
    &\C^{(1)}_{LHL}= -\frac{2}{5 M^6} \, ,&
    &\C^{(2)}_{HLH}= -\frac{3}{20 M^6} \, ,&
    &\C^{(2)}_{LHL}= \frac{2}{15 M^6} \, ,\nn\\
    &\C^{(3)}_{HLH}= -\frac{1}{9 M^6} \, ,&
    &\C^{(3)}_{LHL}= \frac{10-3 \log M^2}{9 M^6} \, ,&
    &\C^{(4)}_{HLH}=  -\frac{2}{9 M^6}\, ,&
    &\C^{(4)}_{LHL}= \frac{20-6 \log M^2}{9 M^6} \, .\nn
\end{align}
Results for classes $\mathcal D^2U^2$, $\mathcal D^2U^3$ and $\mathcal D^2U^4$ are in well-agreement with Ref.~\cite{Ellis:2016enq} \footnote{Note the difference in our notation and that in Ref.~\cite{Ellis:2016enq} where $P_\mu=iD_\mu$ is  treated as the covariant derivative operator.}. The IR-regulator-dependent contributions of these operators are given in the Appendix~\ref{app:A}.
%%%%%%%%%%%%%%%%%%%%%%%%%%%%%%%%%%%%%%%%%%%%%%%%%%
\section{Generalising To Fermionic Effective Action}
\label{subsec:heat-kernel-fermion}
%%%%%%%%%%%%%%%%%%%%%%%%%%%%%%%%%%%%%%%%%%%%%%%%%%

As described in Ref.~\cite{Chakrabortty:2023yke}, the method of HK can be generalized for the fermions by bosonizing the Dirac operator to write it as a second-order strong elliptic operator. This section extends the bosonization procedure for fermions with non-degenerate masses.

We consider a general fermionic Lagrangian with $N$ non-degenerate fermions with scalar and pseudoscalar Yukawa interactions,
\begin{eqnarray}
    \L_{\text{f}} = \overline \Psi_a (i \slashed D\delta_{ab} - M_a\delta_{ab} - \Sigma_{ab}) \Psi_b\;,
\end{eqnarray}
where, the index $a,b$ are the particle index ($12,..,N$), $M_a$ is the mass of the $a^{th}$ particle and the interaction $\Sigma_{ab}$ is given by,
{\small\begin{equation}
    \Sigma_{ab}=S_{ab}+i R_{ab}\ \gamma_5\;.
\end{equation}}
The one-loop fermionic Lagrangian is given by,
\begin{align}
    \L^{(1)}_{\eff}\big\vert_{\text{f}} &= - \, \ln \text{Det} \left[\frac{\delta^2 \L_{\text{f}}}{\delta \overline \Psi \delta \Psi}\right]\;,\nn\\
    &= -\,\ln \text{Det} \left[i \slashed D\delta_{ab} - M_a\delta_{ab} - \Sigma_{ab}\right]\;.
\end{align}
This can be written in the matrix form as,
\begin{eqnarray}
    \frac{\delta^2 \L_{\text{f}}}{\delta \overline \Psi \delta \Psi} = \begin{bmatrix}
        i \slashed D - M_1 - \Sigma_{11} &  -\Sigma_{12} &  -\Sigma_{13}\\
        - \Sigma_{21} &  i \slashed D - M_2 - \Sigma_{22} &  -\Sigma_{23}\\
         - \Sigma_{31} &  -\Sigma_{32} &  i \slashed D - M_3 -\Sigma_{33}\\
    \end{bmatrix}\;.
\end{eqnarray}
Here, for the purpose of demonstrating the bosonization, we have considered three non-degenerate fermions. Using the mass-evenness property of the Dirac operator, i.e. the spectrum of $(i\slashed D)$ and $(-i\slashed D)$ is the same, the one-loop effective Lagrangian can be written as,
\begin{align}\label{eq:fermion_eff}
     \L^{(1)}_{\eff}\big\vert_{\text{f}} &= -\,\ln \text{Det} \left[i \slashed D\delta_{ab} - M_a\delta_{ab} - \Sigma_{ab}\right]\;,\nn\\
     &= -\frac{1}{2}\left(\,\ln \text{Det} \left[i \slashed D\delta_{ab} - M_a\delta_{ab} - \Sigma_{ab}\right] +\,\ln \text{Det} \left[-i \slashed D\delta_{ab} - M_a\delta_{ab} - \Sigma_{ab}\right]\right)\;,\nn\\
     &= -\frac{1}{2}\ln \text{Det}\left[\tilde D_{a}^2\delta_{ab} + M_a^2\delta_{ab} + U^f_{ab}\right]\;.
\end{align}
where the $U^f$ matrix is given by,
\begin{align}
    U^f = \begin{bmatrix}
        Y_{1}+2M_1 \Sigma_{1} +\Sigma^2_{(11)}  & -i(\slashed D \Sigma_{12}) +\Sigma^2_{(12)}  &  -i(\slashed D \Sigma_{13}) +\Sigma^2_{(13)}\\
        -i(\slashed D \Sigma_{21}) +\Sigma^2_{(21)} &  Y_{2}+2M_2 \Sigma_{2}+\Sigma^2_{(22)}  &  -i(\slashed D \Sigma_{23}) +\Sigma^2_{(23)}\\
         -i(\slashed D \Sigma_{31}) +\Sigma^2_{(31)} &  -i(\slashed D \Sigma_{32}) +\Sigma^2_{(32)} &  Y_{3}+2M_3 \Sigma_{3}+\Sigma^2_{(33)} \\
    \end{bmatrix}\;.
\end{align}
We have used the following notations in defining the above $U^f$ matrix,
\begin{align}
    \Sigma_i = \Sigma_{ii},\quad \Sigma^2_{(ij)} = \sum_k \Sigma_{i,k}\Sigma_{k,j},\quad (\slashed D \Sigma_{ij}) =  [\gamma_\mu D_\mu,\Sigma_{ij}],\quad G_{\mu,\nu} = [D_\mu,D_\nu],\nn\\
   \tilde D_{j,\mu} = D_\mu -i\gamma_5\gamma_\mu R_{jj},\quad  Y_j = -\frac{1}{2}\sigma_{\mu\nu}G_{j,\mu\nu} + 4R_{jj}^2-i(\slashed D S_{jj}), \quad \sigma_{\mu\nu} = \frac{1}{2}[\gamma_\mu,\gamma_\nu].
\end{align}
Eq.~\eqref{eq:fermion_eff} has a similar form to that of the scalar one-loop effective Lagrangian. To obtain the effective Lagrangian operators for the fermions, one has to replace the covariant derivatives, field tensors, and interactions in the effective action operators obtained for the scalar case with the corresponding operators of the bosonized fermionic one-loop effective action \cite{Chakrabortty:2023yke} as given below and perform the trace over the Clifford indices,
\begin{align}
    D_{j,\mu}\rightarrow \tilde D_{j,\mu}, \quad F_{j,\mu\nu}\rightarrow G_{\mu\nu}+\Gamma_{j,\mu\nu},\quad U_s\rightarrow U_f\;,
\end{align}
where,\[\Gamma_{j,\mu\nu}= \gamma^5\gamma_\mu (D_\nu R_{jj}) - \gamma^5\gamma_\nu (D_\mu R_{jj}) + 2\sigma_{\mu\nu}R_{jj}^2\;.\]

%%%%%%%%%%%%%%%%%%%%%%%%%%%%%%%%%%%%%%%%%%%%%%%%%%%%%%%%%%%%%%%%%
\section{Algorithm}\label{sec:code}

In the previous sections, we have delineated the path to calculate the effective action for non-degenerate heavy fields using HK prescription and discussed some examples.  Despite the intricate nature of the calculations, it is algorithmic and hence can be automatized using Mathematica \cite{Mathematica}. We are on the verge of developing a Mathematica package to help with the computational challenge. It is accessible `as per request' and is still in its early stages. The ancillary files that are generated from our code are available in \href{https://github.com/shakeel-hep/HK-Effective-Action}{\underline{this link}} and can be used to verify and corroborate our findings. This program uses \texttt{NCAlgebra} \cite{NCAlgebra}, a Mathematica package for non-commutative manipulations. In the section that follows, we outline our program's functionality and map it to the computation of the effective action.

\subsection{Program Description}

Sec.~\ref{sec:heat-kernel-intro} provides a comprehensive overview of the calculation process. Here we connect some of the functions of our yet-to-be-released package with the equations shown in Sec.~\ref{sec:heat-kernel-intro}.
%This program uses \textbf{NCAlgebra} Mathematica package \cite{} for non-commutative manipulations. The various functions used in the program are described below.
\begin{itemize}
\item The first step to initialize the computation is to specify the number of heavy non-degenerate particles. This constructs the matrix $\text{\textbf{A}}'$ (similar to Eq.~\eqref{eq:A_mat}, without the $e^{\pm M^2 t}$). This could be achieved by using the function \textbf{Initialising[N]}.  
\begin{itemize}
    \item[\ding{43}] \textbf{Initialising[}Number of Particles\textbf{]}: Takes the number of particles ($N$) as an input and creates a $N\cross N$ diagonal mass matrix with the diagonal entries $M_i$ ($i=1,2,..N$) and a $N\cross N$ matrix of the operator $\text{\textbf{A}}'$ in $\mathcal A = e^{M^2 t} \text{\textbf{A}}' e^{-M^2 t}$ (Eq.~\eqref{eq:A_mat}).
 \begin{equation}
 \resizebox{.8\hsize}{!}{$
        \text{\textbf{A}}' =\begin{bmatrix}
            D^2 + 2i\, D\cdot p/\sqrt{t} +U_{11}&U_{12}&\cdots&U_{1N}\\
            U_{21}&D^2 + 2i\, D\cdot p/\sqrt{t} +U_{22}\ &\cdots&U_{2N}\\
            .&.&\cdots&.\\
            .&.&\cdots&.\\
            .&.&\cdots&.\\
            U_{N1}&U_{N2}&\cdots&\ D^2 + 2i\, D\cdot p/\sqrt{t} +U_{NN}
        \end{bmatrix}$}.
    \end{equation}
    Here $U_{ij}$ is the interaction between the $i^{\text{th}}$ and $j^{\text{th}}$ particles.
\end{itemize}
    
\item The next step is to use Eq.~\eqref{eq:fn} to obtain the Volterra equation form of $\mathcal{F}(t, \mathcal{A})$ in terms of $f_n (t, \mathcal{A})$. The order $n$ has to be set based on the mass dimension of the operator one wishes to compute.  The detailed files are available in the ``\texttt{.m}" format in \href{https://github.com/shakeel-hep/HK-Effective-Action}{\underline{this link}} and may be simply loaded in a Mathematica notebook \footnote{We will update this repository in future with further details.}. The \textbf{Recursion[}$n$\textbf{]} function, where $n$ determines the series order, performs this step.
\begin{itemize}
    \item[\ding{43}] \textbf{Recursion[}$n$\textbf{]}: Computes $f_n(t,\mathcal A)$ in Eq.~\eqref{eq:fn} up to $n$\textsuperscript{th} order.
\end{itemize}
    
\item As described in section \ref{sec:heat-kernel-intro}, to obtain the effective Lagrangian we have to perform the momentum integral $(p)$ and heat kernel parameter $(t)$ integral, see Eq.~\eqref{eq:final_exp}. The following function carries out the procedure.
\begin{itemize}
    \item[\ding{43}] \textbf{compute[}$term$\textbf{]}: Performs the $p$-integral and $t$-integral of the input, $term$, and gives the output as `` \texttt{Wilson coefficient}$\times$\textbf{func}$[\tr[operator]]$". Here \textbf{func} is just a $head$ that separates the operator from its Wilson coefficient.
\end{itemize}

\item We have some additional functions defined to help categorise the enormous number of terms in different operator classes.
\begin{itemize}
    \item[\ding{43}] \textbf{findterm[}term, Number Of $D$, Number Of $U$\textbf{]}: Used to identify operators of a particular class from a list of operators. An operator class of interest is defined by the number of covariant derivative operators ($D$) and the number of interaction terms ($U$) in it.
\end{itemize}

\end{itemize}

\subsection{Guide to use / Example}
In this subsection, we provide a schematic of the \mathematicanb \; example file. First, the necessary functions and packages are loaded. To install the \href{https://github.com/NCAlgebra}{\texttt{\underline{NCAlgebra}}} we request the user to visit the \href{https://github.com/NCAlgebra}{\underline{website}} for instructions. Once the package is loaded, the interaction matrix and mass matrix are defined by running the \textbf{Initialising} function with the number of particles as an input.
\label{sec:verification}
\begin{mmaCell}[moredefined={Initialising, Number, of, Particles}]{Input}
  Initialising[Number of Particles];
\end{mmaCell}
To compute the $f_n(t,\mathcal A)$ to the desired order ($n$), the following command is run.
\begin{mmaCell}[moredefined={Recursion, n}]{Input}
  Recursion[n];
\end{mmaCell}
The $f_n(t,\mathcal A)$ up to the input order $n$ are saved in the variable $F[k]$ where $k=1,...,N$ and each of the $F[k]$ is a function of $s[k+1]$, the integral variable in Eq.~\eqref{eq:fn}. We want to mention again the \texttt{".m"} files available in \href{https://github.com/shakeel-hep/HK-Effective-Action}{\underline{this link}} named \texttt{"FnMN.m"}, where $n$ denotes the order and $N$ denotes the number of particles (for example, the \texttt{"F6M2.m"} contains the expression $f_6(t, \mathcal{A})$ for 2 particles) can be directly loaded into the notebook file and there is no need to execute the \texttt{Recursion[n]} function. Then to obtain the open derivative operator structures from a particular $f_k(t,\mathcal A)$, first $s[k+1]$ has to be replaced to $t$ in $F[k]$ and $Exp[-M t]$ should be multiplied on the left. Trace of thus obtained matrix has to be given as an input to the \textbf{compute} function. The commands are described below.
\begin{mmaCell}[moredefined={ParallelMap,F, k, compute, trc, M, t, s}]{Input} 
SA[k]=ParallelMap[compute,Expand[Tr[MatrixExp[-M^2 t].F[k]/.\(\pmb{\,}\)s[k+1]\(\pmb{\to}\)t]]];\\
  Export["SAk.m",SA[k]]
\end{mmaCell}

To facilitate the evaluation of the light-heavy mixing contributions, we replace the $\exp[-M^2t]$ with $\exp[-\Lambda^2 t]$, where $\Lambda$ is a $N\cross N$ diagonal matrix with the diagonal entries $\Lambda_i$ ($i=1,2,..N$). This function acts as the IR regulator. To compute the effective Lagrangian operators generated from heavy-light mixed loops, we replace $\Lambda_i$ with $M_i$, $M_i$ being the masses of the heavy particles. The light particle masses $M_j$, which arise from F[k], and $\Lambda_j$'s are then set to be zero sequentially. The following series of commands illustrates this procedure.

\begin{mmaCell}[moredefined={ParallelMap,F,Series,Normal, k, compute, trc,\Lambda, Mi,Mj,i,j, t, s}]{Input} 
SA[k]=ParallelMap[compute,Expand[Tr[MatrixExp[-\(\Lambda\)^2 t].F[k]/.\(\pmb{\,}\)s[k+1]\(\pmb{\to}\)t]]];\\
temp=Series[SA[k]/.\{\(\Lambda\)i\(\pmb{\to}\)Mi\},\{Mj,0,0\}]//Normal;\\
lightHeavy=Series[temp,\{\(\Lambda\)j,0,0\}]//Normal;
\end{mmaCell}

After performing the integral, the operators in open derivative form are obtained. If one wants to collect the operator of class $D^p\,U^q$ from the above expression, one just has to execute the following line.
\begin{mmaCell}[moredefined={ParallelMap,findterm,SA[k]}]{Input}
ParallelMap[findterm[#,p,q]&,SA[k]];
\end{mmaCell}

A point to note is that the operators obtained, from this code, are in their open derivative forms. To reduce them to the usual covariant structures, as mentioned in Sec. \ref{sec:procedure}, one has to consider all the close derivative operators contributing to that particular class with variable coefficients and solve for them by expanding the commutators and comparing them with the operators obtained from the code. This is similar to the method performed in Ref.~\cite{Banerjee:2023iiv}, except now we have to keep track of particle indices on $U$. One can get an idea of all possible closed derivative structures of a particular class from the UOLEA result provided in Ref.~\cite{Banerjee:2023iiv} but it is important to remember that these operators might not be always independent and there can be redundancy in the operators through integration by parts or might be hermitian conjugates \footnote{Operators that are hermitian conjugates of each other have the same Wilson coefficients.}. These redundancies must be removed in order to be able to solve for the coefficients.

%%%%%%%%%%%%%%%%%%%%%%%%%%%%%%%%%

%%%%%%%%%%%%%%%%%%%%%%%%%%%%%%%%%%	
\section{Conclusions}
\label{sec:conclusion}
The Heat-Kernel (HK) provides an algebraic platform to compute the effective action if we can recast the action in terms of a strong elliptic operator. In our earlier papers \cite{Banerjee:2023iiv,Chakrabortty:2023yke}, we have formulated the effective action up to dimension eight after integrating out degenerate heavy scalars and fermions. In the case of the degenerate spectrum, the mass matrix $M$ is proportional to identity and thus commutes with the generic interaction matrix $U$. This allows us to decouple the interaction part of the HK from the free one and rewrite that in terms of a convergent series. But that certainly does not work when we have a non-degenerate spectrum, as $M$ in general does not commute with $U$. Also, the earlier method fails to capture the effects of light-heavy mixing, which emerged at one-loop, in the WCs. In this paper, we have proposed a Heat-Kernel based method to compute the one-loop effective action up to any mass-dimension for non-degenerate scalars and fermions. This method also encapsulates any partial degeneracy in the spectrum, if any, and also the light-heavy mixing effect after ensuring the infrared (IR) safety of the theory. We have computed the most generic (with maximal allowed non-degeneracy) effective action up to dimension six and noted that our results are in good agreement with the previous method. We have considered a toy example of a scalar theory with two non-degenerate scalars and demonstrated the computation of effective operators. We have explicitly shown how we can define an infrared-safe UV theory to start with and compute the light-heavy mixed WCs from the non-degenerate ones. We have also highlighted these contributions into three parts: the finite one, the IR-divergent part, and the contribution that vanishes in the limit light mass is set to zero. In this article, we have outlined the algorithm and Mathematica-based program with an example notebook and supporting files. The algebraic structure of our prescription allows us to construct the algorithm in such a way that can be easily extended to compute effective action up to any mass-dimension in a recursive way using the process discussed in this work.  
%%%%%%%%%%%%%%%%%%%%%%%%%%%%%%%%%%

\appendix
\section{Heavy-Light Mixing Calculation}\label{app:A}
Let us consider the operator $\C_{HLL}\ U_{HL}[D_\mu,U_{LL}][D_\mu,U_{LH}]$ from the operator class $\mathcal D^2 U^3$. Including a regulator and expanding around the light particle mass near $m=0$, up to $\mathcal O(m^2)$, we find the Wilson coefficient as
\begin{align}
    \C_{HLL} =& -\frac{1}{6 \Lambda^2 M^{12}} \Big(M^{10}+M^8 \Big(m^2-3 \Lambda^2\Big)-11 \Lambda^4 M^6-8 M^4 \Big(\Lambda^6+5 \Lambda^4 m^2\Big)-40 \Lambda^6 M^2 m^2\Big)\nn\\&+\Lambda^2 \Big[\Big(M^8+2 M^6 m^2\Big) \log \Big(\frac{M^2}{m^2}\Big)+\Big(\Lambda^2+M^2\Big) \log \Big(\frac{\Lambda^2}{\Lambda^2+M^2}\Big) \Big\{M^6+2 M^4 m^2-\nn\\& 8 \Lambda^4 \Big(M^2+5 m^2\Big)-\Lambda^2 M^2 \Big(7 M^2+20 m^2\Big)\Big\}\Big].
\end{align}

In the above expression, if we set $\Lambda$, different from $m$, there will be terms that mimic the subtraction terms computed in Ref.~\cite{Ellis:2016enq}. According to our proposal where we have introduced an IR-regulator $\Lambda$, we do not need to compute those kinds of terms separately, rather we find the WCs for the local effective operators directly. Here, we express the light-heavy contribution to the WCs into three parts: finite one, the IR-regulator dependent part within [..], and the term that will vanish in the limit light mass to be zero, i.e., $m\rightarrow 0$ within \{..\},  
\begin{align}
    \C_{HLL} =\frac{1}{2 M_1^4}-\left[\frac{\log \frac{\Lambda ^2}{m^2}}{6 M^4}+\frac{1}{6 \Lambda^2 M^2}\right]- \left\{\frac{m^2}{6\Lambda^2 M^4}\right\}+\O{\frac{1}{M^6}}\, .
\end{align}
Below we provide the heavy-light mixing contributions for the classes of operators considered in Sec. \ref{sec:2_paticle_result} in the above mentioned format.

\subsection*{\underline{$\mathcal{O}(\mathcal{D}^2U^2)$}}
\begin{align}
    \C_{LH} = \frac{1}{2M^2} + \O{\frac{1}{M^4}}\, .
\end{align}

\subsection*{\underline{$\mathcal{O}(\mathcal{D}^2U^3)$}}
\begin{align}
    &\C_{LHH} =  -\frac{1}{3M^4} +\O{\frac{1}{M^6}}\, ,& &\C_{LLH} = \frac{5}{2 M^4}-\frac{\log M^2}{M^4}+\left[\frac{\log \Lambda^2}{M^4}\right] +\O{\frac{1}{M^6}}\, ,& \nn\\
    &\C_{HHL} = -\frac{1}{2M^4}+\O{\frac{1}{M^6}}\, ,& &\C_{HLL} =\frac{1}{2 M_1^4}-\left[\frac{\log \frac{\Lambda ^2}{m^2}}{6 M^4}+\frac{1}{6 \Lambda^2 M^2}\right]- \left\{\frac{m^2}{6\Lambda^2 M^4}\right\}+\O{\frac{1}{M^6}}\, .&
\end{align}

\subsection*{\underline{$\mathcal{O}(\mathcal{D}^2U^4)$}}
\begin{align}
    &\C^{(1)}_{LHLH} =\C^{(1)}_{LLHH} = \C^{(1)}_{HLLH} =-\frac{17}{6 M^6} +\frac{\log M^2}{M^6}-\left[\frac{\log \Lambda ^2}{M^6}\right]+\O{\frac{1}{M^8}}\, ,&  \nn\\
    &\C^{(1)}_{HLHL} =\C^{(1)}_{HHLL} =\C^{(1)}_{LHHL} =-\frac{{1}}{M^6} +\frac{1}{M^4}\left[\frac{\log \frac{\Lambda ^2}{m^2}}{3 M^2}+\frac{1}{6 \Lambda ^2}\right]+\left\{\frac{m^2}{3M^6\Lambda^2}\right\}+\O{\frac{1}{M^8}}\, ,&\nn\\
    &\C^{(2)}_{LHLH} = \C^{(2)}_{HLHL} = -\frac{5}{12M^6}+ \frac{1}{M^4}\left[ \frac{\log \frac{\Lambda ^2}{m^2}}{6 M^2}+\frac{1}{12 \Lambda ^2}\right]+ \left\{\frac{m^2}{6M^6\Lambda^2}\right\}+\O{\frac{1}{M^8}}\, ,&  \nn\\
    &\C^{(1)}_{LLLH} = \frac{6}{M^6}-\frac{3 \log M^2}{M^6}-\left[\frac{\log m^2}{M^6}+\frac{1}{2 \Lambda ^2 M^4}+\frac{4 \log \Lambda ^2}{M^6}\right]+\left\{\frac{m^2}{\Lambda ^2 M^6}\right\}+\O{\frac{1}{M^8}}\, ,&\nn\\
    &\C^{(1)}_{HHHL} =\frac{1}{2M^6}+\O{\frac{1}{M^8}}\, ,&\nn\\
    &\C^{(1)}_{HHLH} =\C^{(1)}_{HLHH} = \frac{1}{4M^6}+\O{\frac{1}{M^8}}\, ,&\nn\\
    &\C^{(1)}_{LHLL} =\C^{(1)}_{HLLL} = \C^{(1)}_{LLHL} =\frac{2}{3 M^6}+\left[\frac{\log \frac{m^2}{\Lambda^2}}{3M^6}+\frac{1}{M^4}\left(\frac{1}{12 m^2}-\frac{1}{4 \Lambda ^2}\right)+\frac{1}{12 \Lambda ^4 M^2}\right]\nn\\&\hspace{40mm}-\left\{\frac{m^2}{2M^6 \Lambda ^2}-\frac{m^2}{12 M^4\Lambda ^4}\right\}+\O{\frac{1}{M^8}}\, ,&\nn\\
    &\C^{(2)}_{LLLH}=\C^{(2)}_{HLLL}= \frac{8}{3 M^6}-\frac{\log M^2}{M^6}+\left[\frac{1}{12 m^2 M^4}-\frac{1}{12 \Lambda ^2 M^4}+\frac{1}{12 \Lambda ^4 M^2}+\frac{\log \Lambda ^2}{M^6}\right]\nn\\&\hspace{40mm}-\left\{\frac{m^2}{6 \Lambda ^2 M^6}-\frac{m^2}{12 \Lambda ^4 M^4}\right\}+\O{\frac{1}{M^8}}\, ,&\nn\\
    &\C^{(2)}_{HHHL}=\C^{(2)}_{LHHH}= \frac{5}{12 M^6}+\O{\frac{1}{M^8}}\, ,&\nn\\
    &\C^{(2)}_{LLHH}= -\frac{6}{M^6}+\frac{2 \log M^2}{M^6}-\left[\frac{2 \log \Lambda ^2}{M^6}\right]+\O{\frac{1}{M^8}}\, ,&\nn\\
    &\C^{(2)}_{LHHL}=-\frac{5}{6 M^6} +\left[\frac{1}{6 \Lambda ^2 M^4}+\frac{\log \frac{\Lambda ^2}{m^3}}{3 M^6}\right]+\left\{\frac{m^2}{3 \Lambda ^2 M^6}\right\}+\O{\frac{1}{M^8}}\, .&
\end{align}

\subsection*{\underline{$\mathcal{O}(\mathcal{D}^4U^2)$}}
\begin{align}
    &\C^{(1)}_{LH} = \frac{1}{6M^4}+\O{\frac{1}{M^6}}\, ,&
    &\C^{(2)}_{LH} = \frac{1}{6M^4}+\O{\frac{1}{M^6}}\, ,&\nn\\
    &\C^{(2)}_{HL} = \frac{1}{6M^4}+\O{\frac{1}{M^6}}\, ,&
    &\C^{(4)}_{LH} = \frac{7}{18M^4}-\frac{\log M^2}{6M^4}+\left[\frac{\log \Lambda^2 }{6 M^4}\right]+\O{\frac{1}{M^6}}\, ,& \nn\\
    &\C^{(4)}_{HL} = -\frac{1}{9M^4}+\O{\frac{1}{M^6}}\, ,&
    &\C^{(3)}_{LH} = -\frac{1}{4 M^4}+\frac{1}{ 12M^4}\left[ \log \frac{\Lambda^2}{m^2}+\frac{ M^2}{ \Lambda^2}\right]+\left\{ \frac{m^2}{12M^4\Lambda^2}\right\}+\O{\frac{1}{M^6}}\, ,& \nn\\
    &C^{(3)}_{HL} = \frac{1}{6M^4}+\O{\frac{1}{M^6}}\, .&
\end{align}
%%%%%%%%%%%%%%%%%%%%%%%%%%%%%%%%%%%%%%%%%%%%%%%%%%%%
\subsection*{\underline{$\mathcal{O}(\mathcal{D}^2U^5)$}}
\begin{align}
    &\C^{(1)}_{LHLHL}=-\frac{3 \log M^2-9}{2 M^8}+\left[\frac{1}{12 m^2 M^6}+\frac{3 \log \Lambda ^2}{2 M^8}-\frac{1}{12 \Lambda ^2 M^6}+\frac{1}{24 \Lambda ^4 M^4}\right]\nn\\
    &\hspace{20mm}-\left\{ \frac{m^2}{4 \Lambda ^2 M^8}-\frac{m^2}{12 \Lambda ^4 M^6}\right\}+\O{\frac{1}{M^{10}}}\, ,\nn\\
    &\C^{(1)}_{HLHLH}= -\frac{77-24 \log M^2}{24 M^8}-\left[\frac{\log \Lambda ^2}{M^8}\right]+\O{\frac{1}{M^{10}}}\, ,\nn\\
    &\C^{(1)}_{LHHLL}= -\frac{3-3 \log M^2}{2 M^8}-\left[-\frac{3 \log M^2}{2 M^8}-\frac{1}{6 m^2 M^6}+\frac{3 \log \Lambda ^2}{M^8}-\frac{1}{12 \Lambda ^4 M^4}+\frac{2}{3 \Lambda ^2 M^6}\right]\nn\\
    &\hspace{20mm}-\left\{\frac{2 m^2}{\Lambda ^2 M^8}-\frac{m^2}{6 \Lambda ^4 M^6}\right\}+\O{\frac{1}{M^{10}}}\, ,\nn\\
    &\C^{(1)}_{HLLHH}= -\frac{17-6 \log M^2}{12 M^8}-\left[\frac{\log \Lambda ^2}{2 M^8}\right]+\O{\frac{1}{M^{10}}}\, ,\nn\\
    &\C^{(2)}_{LHLHL}= \frac{7-3 \log M^2}{M^8}-\left[\frac{\log M^2}{M^8}-\frac{4 \log \Lambda ^2}{M^8}-\frac{1}{3 \Lambda ^2 M^6}\right]+\left\{\frac{m^2}{\Lambda ^2 M^8}\right\}+\O{\frac{1}{M^{10}}}\, ,\nn\\
    &\C^{(2)}_{HLHLH}= -\frac{10-3 \log M^2}{3 M^8}-\left[\frac{\log \Lambda ^2}{M^8}\right]+\O{\frac{1}{M^{10}}}\, ,\nn\\
    &\C^{(2)}_{LHHLL}= -\frac{5-3 \log M^2}{M^8}+\left[\frac{1}{6 m^2 M^6}-\frac{5}{6 \Lambda ^2 M^6}+\frac{1}{12 \Lambda ^4 M^4}+\frac{2 \log M^2}{M^8}-\frac{5 \log \Lambda ^2}{M^8}\right]\nn\\&\hspace{20mm}-\left\{\frac{5 m^2}{2 \Lambda ^2 M^8}-\frac{m^2}{6 \Lambda ^4 M^6}\right\}+\O{\frac{1}{M^{10}}}\, ,\nn\\
    &\C^{(2)}_{HLLHH}= \frac{1}{4 M^8}+\O{\frac{1}{M^{10}}}\nn
\end{align}
%%%%%%%%%%%%%%%%%%%%%%%%%%%%%%%%%%%%%%%%%%%%%%%%%%%%
\subsection*{\underline{$\mathcal{O}(\mathcal{D}^4U^3)$}}
\begin{align}
    &\C^{(1)}_{HLH}=  -\frac{1}{20 M^6}+\O{\frac{1}{M^{8}}}\, ,\nn\\
    &\C^{(1)}_{LHL}= -\frac{2}{5 M^6} -\left[\frac{\log \frac{m^2}{\Lambda ^2}}{6 M^6}-\frac{7}{60 \Lambda ^2 M^4}+\frac{1}{30 \Lambda ^4 M^2}+\frac{1}{30 m^2 M^4}\right]-\left\{-\frac{7 m^2}{30 \Lambda ^2 M^6}+\frac{m^2}{30 \Lambda ^4 M^4}\right\}+\O{\frac{1}{M^{8}}}\, ,\nn\\
    &\C^{(2)}_{HLH}= -\frac{3}{20 M^6} +\O{\frac{1}{M^{8}}}\, ,\nn\\
    &\C^{(2)}_{LHL}= \frac{2}{15 M^6}-\left[\frac{1}{60 m^2 M^4}-\frac{1}{60 \Lambda ^2 M^4}+\frac{1}{60 \Lambda ^4 M^2}\right]-\left\{-\frac{m^2}{30 \Lambda ^2 M^6}+\frac{m^2}{60 \Lambda ^4 M^4}\right\}+\O{\frac{1}{M^{8}}} \, ,\nn\\
    &\C^{(3)}_{HLH}= -\frac{1}{9 M^6} +\O{\frac{1}{M^{8}}}\, ,\nn\\
    &\C^{(3)}_{LHL}= -\frac{-10+3 \log M^2}{9 M^6}+\left[\frac{\log \Lambda ^2}{3 M^6}\right]+\O{\frac{1}{M^{8}}} \, ,\nn\\
    &\C^{(4)}_{HLH}=  -\frac{2}{9 M^6}+\O{\frac{1}{M^{8}}}\, ,\nn\\
    &\C^{(4)}_{LHL}= \frac{20-6 \log M^2}{9 M^6}+\left[\frac{2\log \Lambda ^2}{3 M^6}\right]+\O{\frac{1}{M^{8}}}  \, .\nn
\end{align}
%%%%%%%%%%%%%%%%%%%%%%%%%%%%%%%%%%%%%%%%%%%%%%%%%%%%
\section{Generalised UOLEA results up to dimension six}
\label{app:B}
\subsection*{\underline{$\mathcal{O}(U^2)$}}
This class of operators gets contribution from only $f_2(t,\mathcal A)$.

\begin{itemize}
    \item \underline{Covariant operators}
\begin{eqnarray}
    \L_{\eff}[U^2] = \C_{ij}\,\text{tr}(U_{ij}U_{ji})\, .
\end{eqnarray}
    \item \underline{Wilson coefficients}
\begin{align}
    &\C_{ij} = 1-\frac{M_i^2\log M_i^2}{\Delta_{ij}^2}+\frac{M_j^2\log M_j^2}{\Delta_{ij}^2}\, .\nn
\end{align}
\end{itemize}
%%%%%%%%%%%%%%%%%%%%%%%%%%%%%%%%%%%%%%%%%%%%%%%%%%%%
\subsection*{\underline{$\mathcal{O}(U^3)$}}
This class of operators gets contribution from only $f_3(t,\mathcal A)$.

\begin{itemize}
    \item \underline{Covariant operators:}
\begin{eqnarray}
    \L_{\eff}[U^3] = \C_{ijk}\,\text{tr}(U_{ij}U_{jk}U_{ki})\, .
\end{eqnarray}
    \item \underline{Wilson coefficients:}
\begin{align}
    \C_{ijk}=-\frac{M_i^2 \log M_i^2}{\Delta_{ji}^2 \Delta_{ki}^2}-\frac{M_k^2 \log M_k^2}{\Delta_{ik}^2 \Delta_{jk}^2}-\frac{M_j^2 \log M_j^2}{\Delta_{ij}^2\Delta_{kj}^2}\, .\nn
\end{align}
\end{itemize}
%%%%%%%%%%%%%%%%%%%%%%%%%%%%%%%%%%%%%%%%%%%%%%%%%%%%%%%%%%%%%%%%%
\subsection*{\underline{$\mathcal{O}(U^4)$}}
This class of operators gets contribution from only $f_4(t,\mathcal A)$.
\begin{itemize}
    \item \underline{Covariant operators:}
\begin{align}
    \L_\eff[U^4] =& \C_{ijkl}\ \tr(U_{ij}U_{jk}U_{kl}U_{li})\, .
\end{align}
    \item \underline{Wilson coefficients:}
\begin{align}
    \C_{ijkl}=\frac{M_i^2 \log M_i^2}{\Delta_{ji}^2 \Delta_{ki}^2 \Delta_{li}^2}+\frac{M_j^2 \log M_j^2}{\Delta_{ij}^2\Delta_{kj}^2 \Delta_{lj}^2}+\frac{M_k^2 \log M_k^2}{\Delta_{ik}^2 \Delta_{jk}^2 \Delta_{lk}^2}+\frac{M_l^2 \log M_l^2}{\Delta_{il}^2 \Delta_{jl}^2 \Delta_{kl}^2}\, .\nn
\end{align}
\end{itemize}
%%%%%%%%%%%%%%%%%%%%%%%%%%%%%%%%%%%%%%%%%%%%%%%%%%%%%%%%%%%%%%%%%
\subsection*{\underline{$\mathcal{O}(U^5)$}}
This class of operators gets contribution from only $f_5(t,\mathcal A)$.
\begin{itemize}
\item \underline{Covariant operators:}
\begin{eqnarray}
    \L_\eff[U^5] =& \C_{ijklm}\tr(U_{ij}U_{jk}U_{kl}U_{lm}U_{mi})\, .
\end{eqnarray}
\item \underline{Wilson coefficients:}
\begin{align}
    \C_{ijklm}=&-\frac{M_i^2 \log M_i^2}{\Delta_{ji}^2 \Delta_{ki}^2 \Delta_{li}^2\Delta_{mi}^2}-\frac{M_j^2 \log M_j^2}{\Delta_{ij}^2\Delta_{kj}^2 \Delta_{lj}^2\Delta_{mj}^2}-\frac{M_k^2 \log M_k^2}{\Delta_{ik}^2 \Delta_{jk}^2 \Delta_{lk}^2\Delta_{mk}^2}\nn\\&-\frac{M_l^2 \log M_l^2}{\Delta_{il}^2 \Delta_{jl}^2 \Delta_{kl}^2\Delta_{ml}^2}-\frac{M_m^2 \log M_m^2}{\Delta_{im}^2 \Delta_{jm}^2 \Delta_{km}^2\Delta_{lm}^2}\, .\nn
\end{align}
\end{itemize}
%%%%%%%%%%%%%%%%%%%%%%%%%%%%%%%%%%%%%%%%%%%%%%%%%%%%%%%%%%%%%%%%%
\subsection*{\underline{$\mathcal{O}(U^6)$}}
This class of operators gets contribution from only $f_6(t,\mathcal A)$
\begin{itemize}
\item \underline{Covariant operators:}
\begin{align}
    \L_\eff[U^6] =&\, \C_{ijklmn}\tr(U_{ij}U_{jk}U_{kl}U_{lm}U_{mn}U_{ni})\, .
\end{align}
\item \underline{Wilson coefficients:}
\begin{align}
    \C_{ijklmn}= &\frac{M_i^2 \log M_i^2}{\Delta_{ji}^2 \Delta_{ki}^2 \Delta_{li}^2\Delta_{mi}^2\Delta_{ni}^2}+\frac{M_j^2 \log M_j^2}{\Delta_{ij}^2\Delta_{kj}^2 \Delta_{lj}^2\Delta_{mj}^2\Delta_{nj}^2}+\frac{M_k^2 \log M_k^2}{\Delta_{ik}^2 \Delta_{jk}^2 \Delta_{lk}^2\Delta_{mk}^2\Delta_{nk}^2}\nn\\&+\frac{M_l^2 \log M_l^2}{\Delta_{il}^2 \Delta_{jl}^2 \Delta_{kl}^2\Delta_{ml}^2\Delta_{nl}^2}+\frac{M_m^2 \log M_m^2}{\Delta_{im}^2 \Delta_{jm}^2 \Delta_{km}^2\Delta_{lm}^2\Delta_{nm}^2}+\frac{M_n^2 \log M_n^2}{\Delta_{in}^2 \Delta_{jn}^2 \Delta_{kn}^2\Delta_{ln}^2\Delta_{mn}^2}\, .\nn
\end{align}
\end{itemize}
%%%%%%%%%%%%%%%%%%%%%%%%%%%%%%%%%%%%%%%%%%%%%%%%%%%%%%%%%%%%%%%%%
\subsection*{\underline{$\mathcal{O}(\mathcal{D}^2U^2)$}}
These operators get contribution from $f_3(t,\mathcal A)$ and $f_4(t,\mathcal A)$.
\begin{itemize}
\item \underline{Covariant operators:}
\begin{eqnarray}
    \mathcal{L}_{\text{eff}}[\mathcal{D}^2U^2] = \C_{ij}\,\text{tr}([D_{\mu},U_{ij}][D_{\mu},U_{jk}])\, .
\end{eqnarray}
\item \underline{Wilson coefficients:}
\begin{align}
    &\C_{ij} = \cfrac{M_i^2+M_j^2}{2(\Delta_{ij}^2)^2}-\cfrac{M_i^2M_j^2\log\frac{M_i^2}{M_j^2}}{(\Delta_{ij}^2)^3}\, .\nn
\end{align}
\end{itemize}
%%%%%%%%%%%%%%%%%%%%%%%%%%%%%%%%%%%%%%%%%%%%%%%%%%%%%%%%%%%%%%%%%
\subsection*{\underline{$\mathcal{O}(\mathcal{D}^2U^3)$}}
These operators get contribution from $f_4(t,\mathcal A)$ and $f_5(t,\mathcal A)$.
\begin{itemize}
\item \underline{Covariant operators:}
\begin{eqnarray}
    \mathcal{L}_{\text{eff}}[\mathcal{D}^2U^3] =& \C_{ijk} \tr(U_{ij}[D_\mu,U_{jk}][D_\mu,U_{ki}])
    )\, .
\end{eqnarray}
\item \underline{Wilson coefficients:}
\begin{align}\label{eq:gen_D2U3}
    \C_{ijk}=&-\frac{M_i^2 \left(M_j^2+M_k^2\right)+M_k^2 \left(M_j^2-3 M_k^2\right)}{2 (\Delta_{ik}^2)^2 (\Delta_{jk}^2)^2} +\frac{ M_i^2 M_k^2 \log M_i^2}{\Delta_{ij}^2(\Delta_{ki}^2)^3}+\frac{ M_j^2 M_k^2 \log M_j^2}{\Delta_{ij}^2(\Delta_{jk}^2)^3}\nn\\&- \frac{M_k^2 \left(M_i^2 M_j^2 \left(M_i^2+M_j^2-3 M_k^2\right)+M_k^6\right)\log M_k^2}{(\Delta_{ik}^2)^3 (\Delta_{jk}^2)^3}\, .\nn
\end{align}
\end{itemize}
%%%%%%%%%%%%%%%%%%%%%%%%%%%%%%%%%%%%%%%%%%%%%%%%%%%%%%%%%%%%%%%%%
\subsection*{\underline{$\mathcal{O}(\mathcal{D}^2U^4)$}}
These operators get contribution from $f_5(t,\mathcal A)$ and $f_6(t,\mathcal A)$.
\begin{itemize}
\item \underline{Covariant operators:}
\begin{eqnarray}
    \mathcal{L}_{\text{eff}}[\mathcal{D}^2U^4] = \C^{(1)}_{ijkl}\ U_{ij}U_{jk}[D_\mu,U_{kl}][D_\mu,U_{li}] + \C^{(2)}_{ijkl}\ U_{ij}[D_\mu,U_{jk}]U_{kl}[D_\mu,U_{li}] \, .
\end{eqnarray}
\item \underline{Wilson coefficients:}
\begin{align}
    \C^{(1)}_{ijkl} =& \frac{-3 M_l^4 \left(M_i^2+M_j^2+M_k^2\right)+M_l^2 \left(M_i^2 \left(M_j^2+M_k^2\right)+M_j^2 M_k^2\right)+M_i^2 M_j^2 M_k^2+5 M_l^6}{2 (\Delta^2_{il})^2 (\Delta^2_{jl})^2 (\Delta^2_{kl})^2}\nn\\&
    -\frac{M_i^2 M_l^2 \log M_i^2}{\Delta_{ij}^2 \Delta_{ik}^2 (\Delta^2_{il})^3}-\frac{M_j^2 M_l^2 \log M_j^2}{\Delta_{ji}^2 \Delta_{jk}^2 (\Delta^2_{jl})^3}-\frac{M_k^2 M_l^2 \log M_k^2}{\Delta_{ki}^2 \Delta_{kj}^2 (\Delta^2_{kl})^3}\nn\\&
    - \frac{M_l^2 \log M_l^2 \left(M_i^2+M_j^2+M_k^2\right) \left(-3 M_i^2 M_j^2 M_k^2 M_l^2-3 M_l^8\right)}{(\Delta^2_{li})^3 (\Delta^2_{lj})^3 (\Delta^2_{lk})^3}\nn\\&
    -\frac{M_l^8 \log M_l^2  \left(M_i^2 \left(M_j^2+M_k^2\right)+M_i^4+M_j^2 M_k^2+M_j^4+M_k^4\right)}{(\Delta^2_{li})^3 (\Delta^2_{lj})^3 (\Delta^2_{lk})^3}\nn\\&
    -\frac{M_l^2 \log M_l^2 \left(6 M_i^2 M_j^2 M_k^2 M_l^4+M_i^2 M_j^4 M_k^4+M_i^4 M_j^2 M_k^2 \left(M_j^2+M_k^2\right)+3 M_l^{10}\right)}{(\Delta^2_{li})^3 (\Delta^2_{lj})^3 (\Delta^2_{lk})^3}\, ,\nn\\
    \C^{(2)}_{ijkl} =& -\frac{M_i^2 \log M_i^2 \left(M_k^2 M_l^2 \left(-3 M_i^2+M_k^2+M_l^2\right)+M_i^6\right)}{\Delta_{ij}^2 (\Delta^2_{ik})^3 (\Delta^2_{il})^3}\nn\\&
    -\frac{M_j^2 \log M_j^2 \left(M_k^2 M_l^2 \left(-3 M_j^2+M_k^2+M_l^2\right)+M_j^6\right)}{\Delta_{ji}^2 (\Delta^2_{jk})^3 (\Delta^2_{jl})^3}\nn\\&
    -\frac{M_k^2 \log M_k^2 \left(M_i^2 M_j^2 \left(M_i^2+M_j^2-3 M_k^2\right)+M_k^6\right)}{(\Delta^2_{ik})^3 (\Delta^2_{jk})^3 \Delta_{kl}^2}\nn\\&
    -\frac{M_l^2 \log M_l^2 \left(M_i^2 M_j^2 \left(M_i^2+M_j^2-3 M_l^2\right)+M_l^6\right)}{(\Delta^2_{li})^3 (\Delta^2_{lj})^3 \Delta_{lk}^2}\nn\\&
    -\frac{1}{2 (\Delta^2_{ik})^2 (\Delta^2_{il})^2 (\Delta^2_{jk})^2 (\Delta^2_{jl})^2}\Big[-7 M_k^4 M_l^4 \left(M_i^2+M_j^2\right)-7 M_i^4 M_j^4 \left(M_k^2+M_l^2\right)\nn\\&
    \hspace{0.5cm} +3 M_i^4 M_j^4 \left(M_i^2+M_j^2\right)+3 M_k^4 M_l^4 \left(M_k^2+M_l^2\right)-M_j^2 M_k^2 M_l^2 \left(M_j^4+M_k^4+M_l^4\right)\nn\\&\hspace{0.5cm}
    +3 M_i^2 M_j^2 M_k^2 M_l^2 \left(M_i^2+M_j^2+M_k^2+M_l^2\right)+2 M_i^2 M_j^2 \left(M_i^2+M_j^2\right) \left(M_k^4+M_l^4\right)\nn\\&\hspace{0.5cm}
    +2 M_k^2 M_l^2 \left(M_i^4+M_j^4\right) \left(M_k^2+M_l^2\right)-M_i^2 M_j^2 M_k^2 \left(M_i^4+M_j^4+M_k^4\right)\nn\\&\hspace{0.5cm}
    -M_i^2 M_j^2 M_l^2 \left(M_i^4+M_j^4+M_l^4\right)-M_i^2 M_k^2 M_l^2 \left(M_i^4+M_k^4+M_l^4\right)\Big]\, .\nn
\end{align}
\end{itemize}
%%%%%%%%%%%%%%%%%%%%%%%%%%%%%%%%%%%%%%%%%%%%%%%%%%%%%%%%%%%%%%%%%
\subsection*{\underline{$\mathcal{O}(\mathcal{D}^4U^2)$}}
These operators get contribution from $f_4(t,\mathcal A)$, $f_5(t,\mathcal A)$ and $f_6(t,\mathcal A)$.
\begin{itemize}
\item \underline{Covariant operators:}
\begin{align}
    \mathcal{L}_{\text{eff}}[\mathcal{D}^4U^2] =&\, \C^{(1)}_{ij}\,\tr([D_{\mu},[D_{\mu},U_{ij}]][D_{\nu},[D_{\nu},U_{ji}]])+ \C^{(2)}_{ij} \text{tr}([D_{\mu},U_{ij}][D_{\nu},U_{ji}]G_{\nu\mu}) \nn\\
    & + \C^{(3)}_{ij} \text{tr}(U_{ij}U_{ji}G_{\nu\mu}G_{\nu\mu})+\C^{(4)}_{ij} \tr\{(U_{ij}[D_{\mu},U_{ji}]-[D_{\mu},U_{ij}]U_{ji})[D_{\nu},G_{\nu\mu}]\}\, .
\end{align}
\item \underline{Wilson coefficients:}
\begin{align}
    &\C^{(1)}_{ij} =\C^{(2)}_{ij} = \frac{M_i^4+10M_i^2M_j^2+M_j^2}{6(\Delta^2_{ij})^4} -\frac{M_i^2M_j^2(M_i^2+M_j^2)\log\frac{M_i^2}{M_j^2}}{(\Delta^2_{ij})^5}\, ,\nn\\
    &\C^{(3)}_{ij} = \frac{2M_i^4+5M_i^2M_j^2-M_j^2}{12 M_i^2(\Delta^2_{ij})^3} -\frac{M_i^2M_j^2\log\frac{M_i^2}{M_j^2}}{2(\Delta^2_{ij})^4}\, ,\nn\\
    &\C^{(4)}_{ij} = \frac{-2M_i^4-11M_i^2M_j^2+7M_j^2}{18(\Delta^2_{ij})^4} -\frac{M_j^2(-3M_i^4+M_j^4)\log\frac{M_i^2}{M_j^2}}{6(\Delta^2_{ij})^5}\, .\nn
\end{align}
\end{itemize}
%%%%%%%%%%%%%%%%%%%%%%%%%%%%%%%%%%%%%%%%%%%%%%%%%%
%%%%%%%%%%%%%%%%%%%%%%%%%%%
\section*{Acknowledgements}
We acknowledge the fruitful discussions with Sabyasachi Chakraborty, Diptarka Das, Apratim Kabiraj, and Nilay Kundu. 
%%%%%%%%%%%%%%%%%%%%%%%%%%%

%%%%%%%%%%%%%%%%%%%%%%%%%

%%%%%%%%%%%%%%%%%%%%%%
\bibliographystyle{jhep}
\bibliography{ref.bib}
%%%%%%%%%%%%%%%%%%%%%%

\end{document}